\documentclass[11pt]{article}
\usepackage{geometry}
\usepackage{authblk}
\usepackage{url}
\usepackage{cite}
\usepackage{graphicx, caption, subcaption}
\usepackage{amsmath, amssymb, hyperref, multirow, longtable, paracol, booktabs, diagbox, array, makecell, xcolor}

\newtheorem{theorem}{\bf Theorem}[section]

\title{Using covariance of node states to design early warning signals for network dynamics}

\author[ ]{Shilong Yu$^{1}$, Neil G. MacLaren$^{2}$ and Naoki Masuda$^{1, 3, 4}$}

\affil[1]{Department of Mathematics, State University of New York at Buffalo, NY 14260-2900, USA}
\affil[2]{U.S. Army Research Institute for the Behavioral and Social Sciences, VA 22060, USA}
\affil[3]{Institute for Artificial Intelligence and Data Science,
State University of New York at Buffalo, Buffalo, NY 14260-5030, USA}
\affil[4]{Center for Computational Social Science, Kobe University, Kobe, 657-8501, Japan}

\begin{document}
\date{}
\maketitle

\begin{abstract}
Real-life systems often experience regime shifts. An early warning signal (EWS) is a quantity that attempts to anticipate such a regime shift.
Because complex systems of practical interest showing regime shifts are often dynamics on networks, a research interest is to design EWSs for networks, including determining sentinel nodes that are useful for constructing high-quality EWSs. Previous work has shown that the sample variance is a viable EWS including in the case of networks. We explore the use of the sample covariance of two nodes, or sentinel node pairs, for improving EWSs for networks.
We perform analytical calculations in four-node networks and numerical simulations in larger networks to find that the sample covariance and its combination over node pairs is inferior to the sample variance and its combination over nodes; the latter are previously proposed EWSs based on sentinel node selection. The present results support the predominant use of diagonal entries of the covariance matrix (i.e., variance) as opposed to off-diagonal entries in EWS construction.
\end{abstract}

{\flushleft{{\bf Keywords:} early warning signals; network resilience; nonlinear dynamics; stochastic differential equations; bifurcation; critical phenomena}}

\numberwithin{equation}{section}
\section{Introduction}
Many real-life complex systems experience sudden changes, also known as regime shifts, tipping points, or critical transitions, even if the underlying environment or the properties of these systems continuously and gradually change over time. Examples of regime shifts include species extinctions in ecosystems \cite{scheffer2009early, scheffer2015generic}, deforestation \cite{liu2019reduced, wunderling2022recurrent}, onset of epidemic spreading \cite{pastor2015epidemic, southall2021early}, and human diseases \cite{aihara2022dynamical, dablander2023anticipating, van2014critical}.
%
%
It is desirable to be able to anticipate the occurrence of regime shifts so that one can adequately prepare for or prevent catastrophic events, and an early warning signal (EWS) is a quantity that tries to anticipate such regime shifts. Various EWSs have been studied \cite{scheffer2009early, boettiger2013early,
%
%
dakos2012methods, dakos2015resilience, kefi2014early, aihara2022dynamical, maclaren2025applicability}.

A common feature among many complex systems is that they evolve over time and involve interactions between elements within the system. One can model these systems using a network composed of nodes and edges in which the states of the nodes dynamically change over time, that is, using dynamical systems on networks. In network systems, not all nodes are considered to emit good EWSs \cite{boerlijst2013catastrophic, dakos2014critical, dakos2018identifying, weinans2019finding}. Furthermore, one may want to avoid observing each node for constructing an EWS because observation may be costly \cite{biggs2009turning, dakos2010spatial}. These considerations motivate one to select a subset of the nodes, which we call the sentinel nodes, and use them to construct EWSs. Several methods have been proposed to find a good sentinel node set, including
dynamical network biomarker theory \cite{aihara2022dynamical, chen2012detecting},
%
%
those using the dominant eigenvector \cite{boerlijst2013catastrophic, dakos2018identifying,
%
%
patterson2021and}, control theory \cite{aparicio2021structure}, and degree heuristics \cite{dakos2014critical, maclaren2023early}.

A popular EWS given time series observation, no matter whether the system is a network or not, is the sample variance (or the sample standard deviation) over time \cite{biggs2009turning, carpenter2006rising, contamin2009indicators}.
%
%
In the case of networks, a natural EWS is to combine the sample variance (or standard deviation) over a carefully selected set of sentinel nodes, typically taking the average over the sentinel nodes, with the expectation that the averaging suppresses the fluctuation in the sample variance of the individual nodes owing to the law of large numbers, potentially improving the quality of the EWS
\cite{aparicio2021structure, maclaren2023early, masuda2024anticipating}. An unanswered question for this strategy is the use of sample covariances between pairs of nodes for potentially improving network EWSs. If the connectivity between nodes or synchrony of nodes' activity increases towards an impending regime shift,
the sample covariance between nodes may be better at signaling a regime shift than the sample variance.
In fact, some previous studies used the dominant eigenvalue of the sample covariance matrix as EWSs \cite{brock2006variance, suweis2014early, chen2019eigenvalues, dakos2018identifying}. The dominant eigenvalue depends on the covariance between nodes as well as the variance of nodes. Similarly, dynamic network biomarkers are EWSs utilizing the dominant eigenvalue of the Jacobian matrix of the dynamics~\cite{aihara2022dynamical, chen2012detecting}, which depends on all the entries of the Jacobian matrix. However, the dominant eigenvalue only provides one among many possible ways with which to integrate the sample covariance into EWSs.

The aim of this work is to explore EWSs using the sample covariance. Specifically, we ask whether or not averaging the sample covariance over some node pairs improves the quality of EWS compared to averaging the sample variance.
We previously proposed an index $d$ whose maximization enables selecting a sentinel node set for constructing high-quality EWSs when we construct them by averaging the sample variance over the sentinel nodes~\cite{masuda2024anticipating}.
Therefore, we use $d$ to address the present question. We find that the use of the sample covariance does not improve the quality of EWSs, which justifies the use of the sample variance as opposed to covariance for EWSs.
\section{Methods}\label{sec:methods}
\subsection{Dynamical system models}\label{dynamical system models}

We use four dynamical systems on networks. Here, we only explain the coupled double-well dynamics. We explain the other three dynamical systems in section~S1.


The coupled double-well model with dynamical noise is given by 
\begin{equation}\label{coupled double-well}
    \text{d}x_i = \left[-(x_i - r_1)(x_i - r_2)(x_i - r_3) + D\sum_{j = 1}^N A_{ij}x_j + u_i\right]\text{d}t + \sigma_i\text{d}W_i,~~ i \in \{1, \ldots, N\},
\end{equation}
where $W_i$ is a Wiener process independent for different $i$. We set $(r_1, r_2, r_3) = (1, 3, 5)$. We let $u_i = u + \Delta u_i$ and $\sigma_i = \sigma + \Delta \sigma_i$ to represent stress and noise received by node $i$, respectively. We set $\sigma = 0.05$. We vary $u$ or $D$ as a control parameter. When we gradually increase or decrease $u$, we set $D=0.02$ and start from $u = 0$. When we gradually increase $D$, we set $u = 0$ and start from $D = 0$. Finally, when we gradually decrease $D$, we set $u = 0$ and start from $D = 0.1$.
The only difference between the values used here and \cite{masuda2024anticipating} is that we changed $D$ from $0.05$ to $0.02$ because we are using a different set of networks for numerical simulations, and $D = 0.05$ is too large to have sufficiently many control parameter values before experiencing a bifurcation (i.e., regime shift) for some networks. If we do not have sufficiently many control parameter values before a regime shift, we cannot reliably calculate the $d$ value and Kendall's $\tau$ (see section S2 for the definition and usage of $\tau$).

In the case of homogeneous stress and noise, we set $\Delta u_i = \Delta \sigma_i = 0$, $\forall i$.
In the case of heterogeneous stress and noise, we independently draw each $\Delta u_i$, $i\in \{1, \ldots, N\}$, uniformly from $[-0.25, 0.25]$ and each $\Delta \sigma_i$ uniformly from $[-0.9\sigma, 0.9\sigma]$. We use these methods to generate $\Delta u_i$ and $\Delta \sigma_i$ for all the four dynamics models.
Differently from \cite{masuda2024anticipating}, for simplicity, we do not consider the cases in which either $u_i$ or $\sigma_i$ is homogeneous and the other is heterogeneous.

\subsection{Simulations and calculation of early warning signals}\label{calculation of EWS}

We conducted simulations for each combination of dynamical system model (i.e., coupled double-well, mutualistic interaction, gene regulatory, or susceptible-infectious-susceptible (SIS)), network, control parameter (i.e., $u$ or $D$), and direction. We examine two possible simulation directions: ascending and descending. In ascending simulations, we gradually increase the control parameter; in descending simulations, we gradually decrease the control parameter. The control parameter changes until the first bifurcation point in both cases. See section S3 for details of the simulation methods.
A simulation sequence produces a sample covariance matrix $\hat{C}$ at each of the equally distanced values of the control parameter.
At each control parameter value, we compute $\hat{C}$ based on $L$ samples of $(x_1, \ldots, x_N)$ in the equilibrium.

\begin{figure}[htbp]
    \centering           \includegraphics[width=0.4\linewidth]{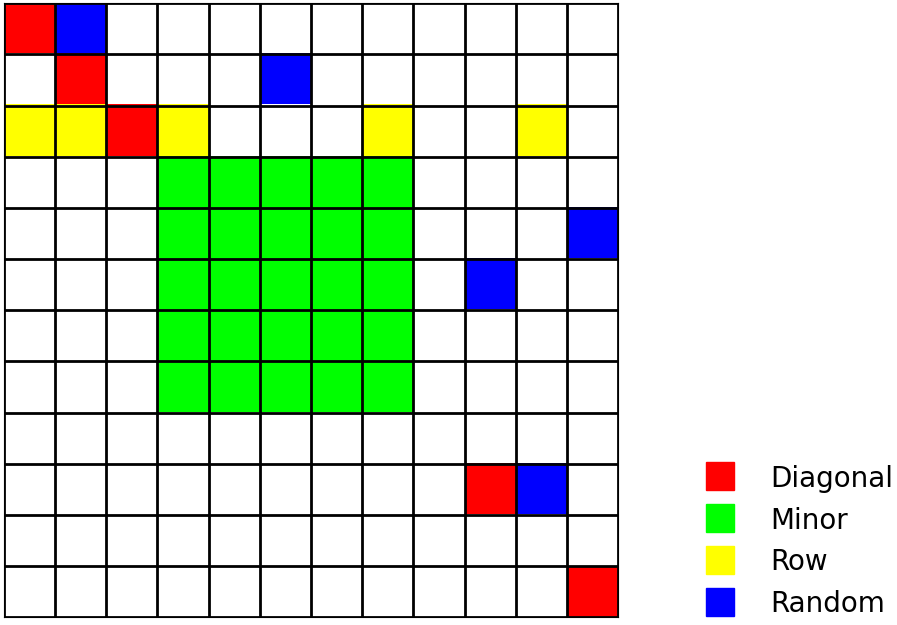}
    \caption{Schematic diagram showing the four methods (diagonal, minor, row, and random) to select entries of the
sample covariance matrix for constructing EWSs.}
\label{figure 2}
\end{figure}

We use four methods to select entries of $\hat{C}$ and calculate an EWS at each control parameter value. These methods only use an $n\times n$ minor matrix of $\hat{C}$, saving monitoring efforts.
In the ``diagonal'' method, we use $n$ diagonal entries of $\hat{C}$,
denoted by $i_1, \ldots, i_n$, selected uniformly at random (shown in red in Fig.~\ref{figure 2}) and calculate the EWS as the average of $\hat{C}_{i_1, i_1}, \ldots, \hat{C}_{i_n, i_n}$ \cite{masuda2024anticipating}. In the ``minor'' method, we use all entries that form a minor matrix (shown in green). Specifically, we sample $n$ numbers from $\{1, \ldots, N\}$ uniformly at random and form a $n \times n$ minor matrix of $\hat{C}$ using these sampled numbers as indices for the rows and columns. We then calculate the EWS as the average of the $n^2$ entries of the minor, of which $n$ are the single-node sample variances and $n^2 - n$ are two-node sample covariances.
In the ``row'' method, we first sample one row of $\hat{C}$ uniformly at random and then use $n$ entries selected uniformly at random from this row (shown in yellow). Then, we calculate the EWS as the average over the $n$ entries.  Finally, in the ``random'' method, we select $n$ entries, which may be on-diagonal or off-diagonal entries, uniformly at random from the $N(N+1)/2$ upper triangular or diagonal entries of $\hat{C}$ (shown in blue) and average them. Although the ``minor'' method uses a larger number of entries of the sample covariance matrix (i.e., $n^2$) than the other three methods (i.e., $n$), the amount of data required for calculating all the four types of EWSs per environment (i.e., control parameter value) is the same, i.e., $L$ samples of $x_i$ from each of the $n$ nodes at equilibrium.

At each control parameter value, we calculate the EWS from $\hat{C}$. Using the EWS values across the control parameter values, we calculate Kendall's $\tau$ between the EWS and the control parameter as a performance measure of the EWS. The value of $\tau$ ranges between $-1$ and $1$. A large $\tau$ in the case of ascending simulations and a small $\tau$ (i.e., close to $-1$) in the case of descending simulations suggest that the EWS performs well. See section S2 for more explanation of $\tau$.

Consider sample covariance matrices $\hat{C}^{(1)}$ and $\hat{C}^{(2)}$, one far from and the other near the first regime shift. See section S3 for how to specifically select the two control parameter values, at which one computes $\hat{C}^{(1)}$ and $\hat{C}^{(2)}$ in our numerical simulations. We select $n$ specific entries according to an entry selection method. Then, we use Eqs.~\eqref{general-mean} and \eqref{general-variance} derived below to calculate the mean and variance of the EWS on the basis of those entries. Note that, for the ``minor'' method, we use $n^2$ instead of $n$ entries of the sample covariance matrix. Then, we calculate 
a previously proposed index $d$ \cite{masuda2024anticipating} given by
	\begin{equation}\label{d}
		d = \frac{|\mu_1 - \mu_2|}{\sqrt{\text{var}_1 + \text{var}_2}},
	\end{equation}
where $\mu_1$ and $\text{var}_1$ are the mean and variance, respectively, of the EWS calculated from $\hat{C}^{(1)}$, and $\mu_2$ and $\text{var}_2$ are those calculated from $\hat{C}^{(2)}$.  For the ``diagonal'' method, it is known that sentinel node sets yielding a large $d$ tend to provide a good EWS in terms of $\tau$
 \cite{masuda2024anticipating}. Note that we only use $d$ to rank sentinel node sets to assess whether or not a sentinel node set would produce a good EWS relative to other sentinel node sets; we do not say that certain $d$ values are large enough to lead to good EWSs.

\section{Results}\label{sec:results}

\subsection{Theory}

We consider an arbitrary network with $N$ nodes and a stochastic dynamical system on it. Let $x_i(t) \in \mathbb{R}$ be the state of the $i$th node at time $t \in \mathbb{R}$. We assume that $x_i(t)$, $i \in \{1, \ldots, N\}$ satisfies the system of stochastic differential equations given by
\begin{equation}\label{n-dim stoch-dy}
	\text{d}\textbf{x}(t) = F(\textbf{x}(t))\text{d}t + B\text{d}W(t),
\end{equation}
where \textbf{x}(t) $= (x_1(t), \ldots, x_N(t))^{\top}$, ${}^{\top}$ denotes the transposition, $F: \mathbb{R}^N \rightarrow \mathbb{R}^N$ is composed of both self dynamics and coupling that involves the adjacency matrix of the network, $B$ is an $N$ by $N$ matrix that is usually a diagonal matrix, and $W(t)$ is an $N$-dimensional vector of independent Wiener processes.
	
	Let $\textbf{x}^* = (x_1^*, \ldots, x_N^*)^\top$ be the equilibrium of the dynamics given by Eq.~\eqref{n-dim stoch-dy} without the noise term, or the solution of $F(\textbf{x}^*) = \textbf{0}$. By linearizing Eq.~\eqref{n-dim stoch-dy} around $\textbf{x}(t) = \textbf{x}^*$, we obtain
	\begin{equation}
		\text{d}\textbf{z}(t) = -A\textbf{z}(t)\text{d}t + B\text{d}W(t),
	\end{equation}
	where $\textbf{z}(t) = \textbf{x}(t) - \textbf{x}^*$, and $A$ is an $N$ by $N$ matrix such that $-A$  is the Jacobian matrix of $F$ at $\textbf{x}(t) = \textbf{x}^*$. Equilibrium $\textbf{x}^*$ is asymptotically stable if and only if $A$ is positive definite.
	
	Let $C = (C_{ij})$ be the $ N \times N$ covariance matrix; $C_{ij}$ represents the covariance of $x_i(t)$ and $x_j(t)$ at equilibrium. One can obtain $C$ from the Lyapunov equation given by
	\begin{equation}
		AC + CA^\top = BB^\top.
	\end{equation}
Matrix $C$ is unique if $A$ is positive definite.

Prior work considered the sample variance of $x_i(t)$ and its average over a selected node set (i.e., the average over a selected diagonal entries of $C$) as EWSs \cite{maclaren2023early, masuda2024anticipating}. In this work, we consider using sample covariance between nodes (i.e., off-diagonal entries of the covariance matrix) with the aim of improving EWSs.
	
Suppose that we observe $L$ samples of $x_i(t)$ for each node $i$ at equilibrium. Then, the unbiased sample covariance between nodes $i$ and $j$ is
	\begin{equation}
		\hat{C}_{i,j} = \frac{1}{L-1}\sum_{\ell = 1}^L (x_{i,\ell} - \overline{x}_i)(x_{j,\ell}-\overline{x}_j),
	\end{equation}
where $x_{i,\ell}$ is the $\ell$th sample of $x_i(t)$ at equilibrium, and $\overline{x}_i := \sum_{\ell=1}^L x_{i,\ell} / L$. We assume that $\textbf{x}_{\ell} := (x_{1, \ell}, \ldots, x_{N, \ell})$ is i.i.d. Note that $z_{i, \ell} := x_{i, \ell} - x^*_i$ satisfies $\mathbb{E}[z_{i,\ell}] = 0$, $\forall i \in \{1, \cdots, N\}$, $\forall \ell \in \{1, \ldots, L\}$, where $\mathbb{E}$ denotes the expectation. Then, we obtain
	\begin{equation}\label{Expectation Cov_{i,j}}
		\mathbb{E}[\hat{C}_{i,j}] = C_{ij}
	\end{equation}
	and 
	\begin{equation}\label{variance of Cov_{i,j}}
		\text{var}(\hat{C}_{i,j}) = \frac{1}{L-1}\left[ C_{ii}C_{jj} + (C_{ij})^2\right],
	\end{equation}
where $\text{var}$ represents the variance.	
More generally, if we use the average of $n$ entries of the sample covariance matrix, $\hat{C}_{i_1,j_1}, \ldots, \hat{C}_{i_n,j_n}$, as EWS, then we obtain
	\begin{equation}\label{general-mean}
		\mathbb{E}\left[\frac{1}{n}\sum_{m = 1}^n \hat{C}_{i_m, j_m}\right] = \frac{1}{n}\sum_{m= 1}^n C_{i_m, j_m}
	\end{equation}
	and 
	\begin{equation}\label{general-variance}
		\text{var}\left(\frac{1}{n}\sum_{m = 1}^n \hat{C}_{i_m, j_m}\right) = \frac{1}{n^2(L-1)}\left(\sum_{m,m'=1}^n C_{i_m,i_{m'}}C_{j_m,j_{m'}} + \sum_{m,m'=1}^n C_{i_m,j_{m'}}C_{i_{m'},j_m}\right).
	\end{equation}
See sections S4 and S5 for the derivation of these results.
			
\subsection{Nonlinear dynamics on networks with four nodes} 

\begin{figure}[ht]
    \includegraphics[width=\textwidth]{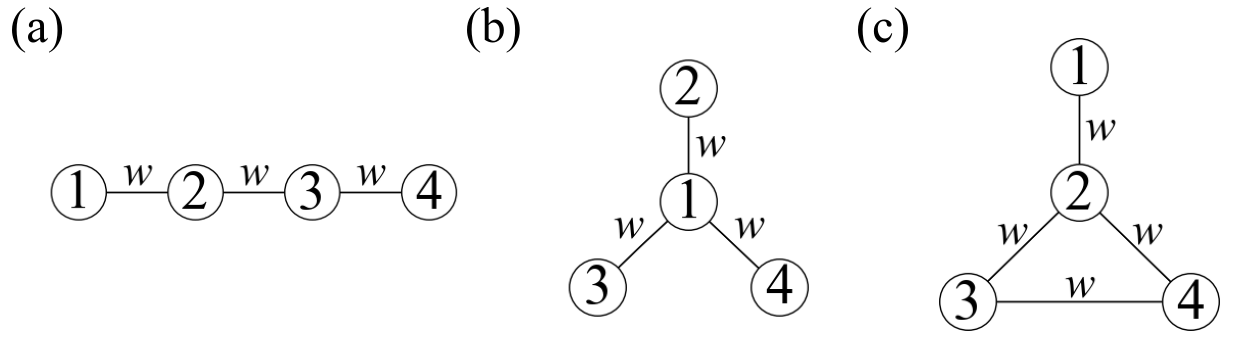}
    \caption{Four-node networks. (a) Chain. (b) Star. (c) Lollipop.}
    \label{fig:mainfigure}
\end{figure}

In this section, we investigate whether the use of sample covariance
%
%
improves the quality of the EWS relative to sample variance in small networks. We consider three networks: the chain, star, and lollipop graphs, shown in Fig.~\ref{fig:mainfigure}. For each network, we calculate the equilibrium without noise, the bifurcation point, the Jacobian, and the covariance matrix by using the Lyapunov equation. We then identify the set of the covariance matrix entries that produces the largest $d$ value. For all three networks, we find that the majority of the maximizers of $d$ only use the diagonal entries of the covariance matrix, i.e., single-node variance.

In our prior work \cite{masuda2024anticipating}, we analyzed networks with $N=2$ and $N=3$ nodes to obtain analytical insights into the quality of the EWS constructed as averages of sample variance (as opposed to covariance) over some nodes. However, there are relatively few off-diagonal entries of $\hat{C}$ with $N=2$ or $N=3$. Therefore, here we consider the chain, star, and lollipop graphs composed of $N=4$ nodes of identical nonlinear dynamics with potentially different dynamical noise strengths and connectivity at each node. We chose these networks because they have some symmetry to be exploited for analytical computations.
We assume that the networks are undirected. For all three networks, dynamical noise would make the bifurcation occur earlier than when there is no noise. However, we are not particularly interested in noise-induced transitions; we instead focus on the control parameter range  approaching and before the saddle-node bifurcations and look for high-quality EWSs. We also assume small noise because the dynamics under large noise would mostly stay far from the equilibrium such that we cannot justify linearization of the stochastic dynamical system.

\subsubsection{Chain}

We first consider the chain network (see Fig.~\ref{fig:mainfigure}(a)).
The system of stochastic differential equations is given by
\begin{align}
        \text{d}x_1 &= [f(x_1) + w(x_2 + 1)]\text{d}t + \sigma_1 \text{d}W_1,
		\\ \text{d}x_2 &= [f(x_2) + w(x_1 + 1) + w(x_3 + 1)]\text{d}t + \sigma_2 \text{d}W_2,
		\\ \text{d}x_3  &= [f(x_3) + w(x_2 + 1) + w(x_4 + 1)]\text{d}t + \sigma_3 \text{d}W_3,
		\\ \text{d}x_4 &= [f(x_4) + w(x_3 + 1)]\text{d}t +\sigma_4 \text{d}W_4,
\end{align}
where $w$ ($>0$) represents the coupling strength (i.e., edge weight). As in \cite{masuda2024anticipating}, we assume $f(x) = r + x^2$ because this $f$ provides the topological normal form of the saddle-node bifurcation when without the coupling term and dynamical noise. Therefore, as $r$ starts with a sufficiently negative value and gradually increases, each node, if isolated, undergoes a saddle-node bifurcation. We further assume that each $x_i$, $i \in \{1, \ldots, 4\}$ satisfies $-1 \leq x_i < 0$. Thus, terms such as $w(x_2 + 1)$, which is guaranteed to be nonnegative, indicate that $x_1$ receives positive input from $x_2$ because nodes $1$ and $2$ are adjacent in the chain network. We use the same $f$ in the other two four-node networks.
	
The equilibrium in the absence of noise satisfies
 \begin{align}
		r + (x_1^*)^2 + w(x_2^* + 1) &= 0, 	\label{chain-1}
		\\ r + (x_2^*)^2 + w(x_1^* + 1) + w(x_3^* + 1) &= 0,         \label{chain-3}
        \\ r + (x_3^*)^2 + w(x_2^* + 1) + w(x_4^* + 1) &= 0,
        \label{chain-2}
		\\ r + (x_4^*)^2 + w(x_3^* + 1) &= 0.
\end{align}
We consider symmetric solutions satisfying $x_1^* = x_4^*$ and $x_2^* = x_3^*$. 
Then, Eq.~\eqref{chain-3} yields
\begin{equation}
r + (x_2^*)^2 + w(x_1^* + 1) + w(x_2^* + 1) = 0.
\label{chain-5new}
\end{equation}
Equations~\eqref{chain-1} and \eqref{chain-5new} define parabolas in the $(x_1^*, x_2^*)$ space. As in prior work \cite{masuda2024anticipating}, we set $w = 0.05$. As we gradually increase the control parameter $r$ from $-1$, the two parabolas initially have four intersections, and then two, and then no intersection. We denote by $r_c'$ the value of $r$ at which the number of intersections changes from four to two, and by $r_c$ the value of $r$ at which the number of intersections changes from two to zero. We have $-1 < r_c'< r_c < 0$. Numerically, $r_c' \approx -0.112$ and $r_c \approx -0.089$. The sign-flipped Jacobian is
	\begin{equation}
		A = \begin{pmatrix}
			-2 x_1^* & -w & 0 & 0 \\
			-w & -2 x_2^* & -w & 0 \\
			0 & -w & -2 x_2^* & -w \\
			0 & 0 & -w & -2 x_1^*
		\end{pmatrix}.
	\end{equation}

To analytically compute the uncertainty of the EWS, we further assume that $\sigma_1 = \sigma_4$ and $\sigma_2 = \sigma_3$. Under these assumptions, $x_1$ and $x_4$ are statistically the same, so are $x_2$ and $x_3$.
We then have $B = \text{diag}(\sigma_1, \sigma_2, \sigma_2, \sigma_1)$, where diag$(a_1, \ldots, a_N)$ denotes the $N$ by $N$ diagonal matrix with $a_i$ in the $(i, i)$ entry. We find that, out of the four intersections (i.e., equilibria), only one is stable; matrix $A$ evaluated at this equilibrium only has negative eigenvalues. Therefore, we evaluate $A$ at this stable equilibrium $\textbf{x}^* = (x_1^*, x_2^*, x_2^*, x_1^*)$, where $x_1^* < 0$ and $x_2^* < 0$. Using the Lyapunov equation, we explicitly obtain matrix $C$ (see section S6 for details). 

We fix $\sigma_2 = 0.1$ and vary $\sigma_1$. We let $L = 100$ and $w = 0.05$ for all three small networks. We compute
$d$ using Eq.~\eqref{d}, where we compute $\mu_1$ and $\text{var}_1$ at $r = -0.3$, and $\mu_2$ and $\text{var}_2$ at $r = -0.12$. We do so because $r = 0.3$ is far from $r_c'$ and $r = -0.12$ is close to $r_c'$. We expect that an EWS that produces large $d$ is better at anticipating the impending saddle-node bifurcation.
To find good EWSs in terms of $d$, we maximize $d$ by allowing the EWS to use one to four entries of the covariance matrix $C$. For each number of allowed entries, $n$, we considered all possible combinations of the entries of $C$ and calculated $d$ in each case. Then, we recorded the maximum possible $d$ value for each $n$.

We show the results in Table \ref{table: chain} for four values of $\sigma_1$; because we fixed the $\sigma_2$ value, the $\sigma_1$ value controls how much dynamic noise the two outer nodes (i.e., nodes $1$ and $4$) in the chain receive relative to the two inner nodes (i.e., nodes $2$ and $3$).
Table \ref{table: chain} indicates that, if $\sigma_1 = 0.004$, then 
the entries maximizing $d$ avoid the inner nodes except for $n=4$, because $\sigma_2$ is much larger than $\sigma_1$.
Furthermore, the best result (i.e., largest $d$) across the $n$ values is obtained only by using $n=2$ on-diagonal entries of the covariance matrix. Additionally using the off-diagonal entries (with $n=3$ and $n=4$) is detrimental. If $\sigma_1 = 0.02$, although the outer nodes still receive much less noise than the inner nodes, the entries that help maximize $d$ are the variance of all the nodes, i.e., $C_{11}$, $C_{22}$, $C_{33}$, and $C_{44}$. Off-diagonal entries of $C$ are not used, and the maximized $d$ value is similar among $n=2$, $3$, and $4$. The results for $\sigma_1 = 0.1$ are qualitatively the same as those for $\sigma_1 = 0.02$. Finally, when $\sigma_1 = 0.5$, the outer nodes receive far more noise than the inner nodes. In this situation, the optimized EWSs
avoid the outer nodes when $n = 1$, $2$, and $3$ because their variances are large. Although just one entry selection, i.e., that with $n=3$, includes an off-diagonal entry (i.e., $C_{23}$), the largest $d$ value across the $n$ values occurs when $n=2$ and is realized by $C_{22}$ and $C_{33}$, not using off-diagonal entries. Overall, these results suggest that using off-diagonal entries of the covariance matrix does not help the quality of EWSs. In the following sections, we will further test this hypothesis by analyzing two other networks with $N=4$ and then running larger-scale numerical simulations.

\begin{longtable}{|>{\centering\arraybackslash}m{0.02\textwidth}|%
                  >{\centering\arraybackslash}m{0.19\textwidth}|%
                  >{\centering\arraybackslash}m{0.19\textwidth}|%
                  >{\centering\arraybackslash}m{0.19\textwidth}|%
                  >{\centering\arraybackslash}m{0.19\textwidth}|}
\caption{Maximum $d$ values and the corresponding combinations of the entries of the covariance matrix, $C$, in the case of the chain network. We set $\sigma_2 = 0.1$. The number of node pairs used for computing the EWS is denoted by $n$. We considered all possible combinations of node pairs. Each entry of the table shows the one that maximizes $d$ for the given $n$ and $\sigma_1$ values. We similarly obtain tables \ref{table: star} and \ref{table: triangle and line}.} 
\label{table: chain}
\endfirsthead
\endhead
\endfoot
\hline
\diagbox[width=2.0em, height=2.2em, innerleftsep=0.4em, innerrightsep=0.2em]{$n$}{$\sigma_1$} & \makecell{\vspace{-0.6em}0.004} & \makecell{\vspace{-0.6em}0.02} & \makecell{\vspace{-0.6em}0.1} & \makecell{\vspace{-0.6em}0.5} \\ \hline
\makecell{\vspace{-0.6em}1} & \makecell{$C_{11}$ \\ $d = 5.54$} & \makecell{$C_{22}$ \\ $d = 3.80$} & \makecell{$C_{22}$ \\ $d = 3.81$} & \makecell{$C_{22}$ \\ $d = 4.04$} \\ \hline
\makecell{\vspace{-0.6em}2} & \makecell{ $C_{11}, C_{44}$ \\ $d = 7.77$ } & \makecell{$C_{22}, C_{33}$ \\ $d = 5.34$} & \makecell{$C_{22}, C_{33}$ \\ $d = 5.36$} & \makecell{$C_{22}, C_{33}$ \\ $d = 5.68$} \\ 
\hline
\makecell{\vspace{-0.6em}3} & \makecell{$C_{11}, C_{14}, C_{44}$ \\ $d = 6.84$} & \makecell{$C_{11}, C_{22}, C_{33}$ \\ $d = 5.41$} & \makecell{$C_{11}, C_{22}, C_{33}$ \\ $d = 6.06$} & \makecell{$C_{22}, C_{23}, C_{33}$ \\ $d = 5.07$} \\ 
\hline
\makecell{\vspace{-0.6em}4} & \makecell{ $C_{11}, C_{12}, C_{34}, C_{44}$ \\ $d = 6.31$ } & \makecell{$C_{11}, C_{22}, C_{33}, C_{44}$ \\ $d = 5.48$} & \makecell{$C_{11}, C_{22}, C_{33}, C_{44}$ \\ $d = 6.09$} & \makecell{$C_{11}, C_{22}, C_{33}, C_{44}$ \\ $d = 4.34$} \\ \hline 
\end{longtable}

\subsubsection{Star}
	
In this section, we consider the same nonlinear dynamics on the star graph (see Fig.~\ref{fig:mainfigure}(b)). To exploit the symmetry of the graph, we assume $\sigma_2 = \sigma_3 = \sigma_4$.
The derivation of the entries of $C$ is similar to the case of the chain network and shown in section S6. We set $\sigma_2 = 0.1$ and vary $\sigma_1$.

\begin{longtable}{|>{\centering\arraybackslash}m{0.02\textwidth}|
                  >{\centering\arraybackslash}m{0.19\textwidth}|%
                  >{\centering\arraybackslash}m{0.19\textwidth}|%
                  >{\centering\arraybackslash}m{0.19\textwidth}|%
                  >{\centering\arraybackslash}m{0.19\textwidth}|}
\caption{Maximum $d$ and the corresponding combinations of the entries of $C$ in the case of the star graph. We set $\sigma_2 = 0.1$. 
%
%
}  
\label{table: star}
\endfirsthead
\endhead
\endfoot
\hline
\diagbox[width=2.0em, height=2.2em, innerleftsep=0.4em, innerrightsep=0.1em]{$n$}{$\sigma_1$} & \makecell{\vspace{-0.6em} 0.004} & \makecell{\vspace{-0.6em}0.02} & \makecell{\vspace{-0.6em}0.1} & \makecell{\vspace{-0.6em}0.5} \\ \hline
\makecell{\vspace{-0.6em}1} & \makecell{$C_{11}$ \\ $d = 4.16$} & \makecell{$C_{11}$ \\ $d = 2.51$} & \makecell{$C_{11}$ \\ $d = 2.24$} & \makecell{$C_{11}$ \\ $d = 2.23$} \\ \hline
\makecell{\vspace{-0.6em}2} & \makecell{ $C_{22}, C_{33}$ \\ $d = 2.41$ } & \makecell{$C_{22}, C_{33}$ \\ $d = 2.41$} & \makecell{$C_{11}, C_{22}$ \\ $d = 2.81$} & \makecell{$C_{22}, C_{33}$ \\ $d = 2.60$} \\ 
\hline
\makecell{\vspace{-0.6em}3} & \makecell{$C_{22}, C_{33}, C_{44}$ \\ $d = 2.95$} & \makecell{$C_{22}, C_{33}, C_{44}$ \\ $d = 2.95$} & \makecell{$C_{11}, C_{22}, C_{33}$ \\ $d = 3.28$} & \makecell{$C_{22}, C_{33}, C_{44}$ \\ $d = 3.18$} \\ 
\hline
\makecell{\vspace{-0.6em}4} & \makecell{ $C_{22}, C_{23}, C_{33}, C_{44}$ \\ $d = 2.97$ } & \makecell{$C_{11}, C_{22}, C_{33}, C_{44}$ \\ $d = 3.02$} & \makecell{$C_{11}, C_{22}, C_{33}, C_{44}$ \\ $d = 3.70$} & \makecell{$C_{22}, C_{23}, C_{33}, C_{44}$ \\ $d = 2.99$} \\ \hline 
\end{longtable}

We show the maximized $d$ value and the corresponding combination of the entries of the covariance matrix in Table \ref{table: star}.
We find that, when $\sigma_1 = 0.004$, the largest $d$ value occurs when we use only the variance of the hub node (i.e., $C_{11}$). This is expected because the hub node receives much smaller noise than the leaf nodes. For $\sigma_1 = 0.02$ and $\sigma_1 = 0.1$, the largest $d$ value occurs when one uses all the diagonal entries of $C$. Off-diagonal entries are not selected for any $n \in \{1, \ldots, 4 \}$. When $\sigma_1= 0.5$, the maximizer of $d$ for each $n$ still only uses on-diagonal entries of $C$ except for $n=4$. The maximum $d$ across the values of $n$ occurs when one uses the variances of the three leaf nodes (i.e., $C_{22}$, $C_{33}$, and $C_{44}$). Overall, the results for the star graph continue to support the hypothesis that using off-diagonal entries does not help the quality of EWSs.

\subsubsection{Lollipop}
	
Consider the lollipop graph shown in Fig.~\ref{fig:mainfigure}(c).
To exploit the symmetry of the graph, we assume $\sigma_3 = \sigma_4$. The derivation of the entries of $C$ is similar to that in the case of the chain and star graphs and shown in section S6.
    
We show the EWSs maximizing $d$ for different values of $\sigma_1$ and $\sigma_2$, with $\sigma_3 = 0.1$ fixed,
in Table~\ref{table: triangle and line}. Because we independently vary $\sigma_1$ and $\sigma_2$, for each given pair of $\sigma_1$ and $\sigma_2$, Table~\ref{table: triangle and line} only shows the set of entries that maximizes $d$ over the $n$ values as well as over the entry selection.
We find that the entry combinations that maximize $d$ are on the diagonal for 12 out of the 16 combinations of $\sigma_1$ and $\sigma_2$.

\begin{longtable}{|>{\centering\arraybackslash}m{0.06\textwidth}|%
                  >{\centering\arraybackslash}m{0.19\textwidth}|%
                  >{\centering\arraybackslash}m{0.19\textwidth}|%
                  >{\centering\arraybackslash}m{0.19\textwidth}|%
                  >{\centering\arraybackslash}m{0.19\textwidth}|}
\caption{Maximum $d$ and the corresponding combinations of the entries of $C$ in the case of the lollipop graph. We set $\sigma_3 = 0.1$.}  
\label{table: triangle and line}
\endfirsthead
\endhead
\endfoot
\hline
\diagbox[width=3.7em, height=2.2em, innerleftsep=0.5em, innerrightsep=0.5em]{$\sigma_1$}{$\sigma_2$} & \makecell{\vspace{-0.6em}0.004} & \makecell{\vspace{-0.6em}0.02} & \makecell{\vspace{-0.6em}0.1} & \makecell{\vspace{-0.6em}0.5} \\ \hline
\makecell{\vspace{-0.6em}0.004} & \makecell{$C_{11}, C_{22}$ \\ $d = 5.23$} & \makecell{$C_{22}, C_{23}, C_{33}, C_{44}$ \\ $d = 3.44$} & \makecell{$C_{12}, C_{22}, C_{33}, C_{44}$ \\ $d = 4.32$} & \makecell{$C_{11}, C_{12}, C_{33}, C_{44}$ \\ $d = 4.95$} \\ \hline
\makecell{\vspace{-0.6em}0.02} & \makecell{ $C_{22}$ \\ $d = 4.93$ } & \makecell{$C_{11}, C_{22}$ \\ $d = 3.63$} & \makecell{$C_{11}, C_{22}, C_{33}, C_{44}$ \\ $d = 4.33$} & \makecell{$C_{11}, C_{12}, C_{33}, C_{44}$ \\ $d = 4.65$} \\ 
\hline
\makecell{\vspace{-0.6em}0.1} & \makecell{$C_{22}$ \\ $d = 5.02$} & \makecell{$C_{11}, C_{22}, C_{33}, C_{44}$ \\ $d = 3.95$} & \makecell{$C_{11}, C_{22}, C_{33}, C_{44}$ \\ $d = 4.72$} & \makecell{$C_{11}, C_{33}, C_{44}$ \\ $d = 4.19$} \\ 
\hline
\makecell{\vspace{-0.6em}0.5} & \makecell{ $C_{22}$ \\ $d = 5.27$ } & \makecell{$C_{22}$ \\ $d = 4.24$} & \makecell{$C_{22}, C_{33}, C_{44}$ \\ $d = 4.36$} & \makecell{$C_{33}, C_{44}$ \\ $d = 3.59$} \\ \hline     
\end{longtable}

In three out of the 16 combinations, simply using all the diagonal entries maximizes $d$. The amount of noise given to each node tended to be relatively similar to each other in these cases. In four other combinations, only using the sample variance of node 2 maximizes $d$. In these cases, node 2 tended to receive much smaller noise than the other three nodes. In the other five combinations for which the maximizer of $d$ only uses the diagonal entries of $C$, either two or three nodes are used, and the association between the selected nodes and the strength of the noise, $\sigma_i$, is weaker.
%
%
In conclusion, the results of the lollipop graph continue to support our claim that off-diagonal entries of $C$ do not improve the quality of EWSs in a majority of cases.
	
\subsection{Numerical results}\label{numerical results}

We ran numerical simulations on four dynamical system models on networks (i.e., coupled double-well, mutualistic interaction, gene regulatory, and susceptible-infectious-susceptible (SIS) models) to investigate whether or not the use of sample covariance improves the quality of EWSs in larger networks without particular symmetry. In sum, we used various methods of selecting sentinel node sets in order to construct EWSs and found that the ``diagonal'' method (i.e., the one without using covariance) produces the best EWS.

We compare four methods of selecting $n$ entries from the sample covariance matrix to construct EWSs, i.e., 
the ``diagonal'', ``minor'', ``row'', and ``random'' methods described in section~\ref{calculation of EWS}
(see Fig.~\ref{figure 2} for a schematic). The ``minor'' method in fact uses $n^2$ entries of the covariance matrix. All but the ``diagonal'' method use sample covariance in general.

We proved the following properties for these sentinel node sets assuming that all entries of $C$ are positive. We show their proofs in section S7.
\begin{theorem}
    \label{thm:theorem 1}
    If $n = 2$, the coefficient of variation (CV; the standard deviation divided by the mean) of the EWS obtained by the ``diagonal" method is always less than or equal to that obtained by the ``row" method.
\end{theorem}

\begin{theorem}
    \label{thm:theorem 2}
    For any $n$, the CV of the EWS obtained by the ``diagonal" method is always less than or equal to that obtained by the ``minor" method.
\end{theorem}
The CV quantifies statistical fluctuations of the EWS relative to its size in a different manner from $d$. A small CV is likely to be associated with a better quality of the EWS because it implies a small relative uncertainty. Therefore, these theorems suggest that the ``diagonal'' method may perform better than the ``row'' and ``minor'' methods under wide conditions.
We set $n=5$ in all the following simulations.

\begin{figure}
    \centering 
     \includegraphics[width=0.7\linewidth]{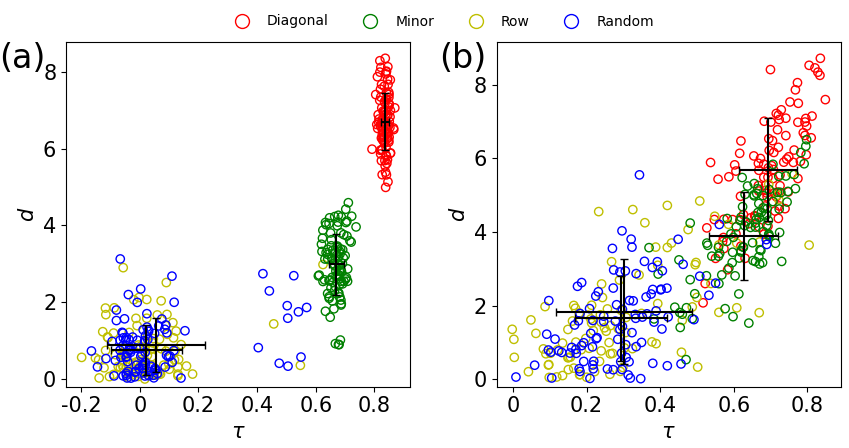}
    \caption{Demonstration of four node set selection methods. (a) Performance of EWSs of different types. A circle
represents one EWS. The
solid error bars show the mean and standard deviation calculated on the basis of the $100$ EWSs of the same type. We used the coupled double-well dynamics on the BA network with 100 nodes with homogeneous stress and noise. The control parameter is $u$, which we gradually increased. (b) Performance of EWSs for the SIS dynamics on the dolphin network with $N=62$ nodes with heterogeneous noise. The control parameter is $D$, which we gradually increased.}
    \label{figure 3}
\end{figure}

As an example, we consider the coupled double-well dynamics (see Eq.~\eqref{coupled double-well}) on a network generated by the Barab\'{a}si-Albert (BA) model \cite{barabasi1999emergence} with $N=100$ nodes. See section S8 for a detailed explanation of the network. We use $u$ as the control parameter and linearly increase it.
Figure~\ref{figure 3}(a) shows the results for the coupled double-well dynamics on the generated BA network. A circle represents an EWS, which is an unweighted average of $n$ entries of the sample covariance matrix. (In the case of the ``minor'' method, it is an unweighted average of $n^2$ entries of the minor matrix.) For each EWS, we calculated $d$ and the Kendall's $\tau$. The colors of the circles correspond to the method of selecting entries. We generated $100$ EWSs of each type, which differ in how we randomly select $n$ entries. Therefore, there are $100$ circles for each color in Fig.~\ref{figure 3}(a). The horizontal and vertical error bars show the mean and standard deviation based on the $100$ samples for each color.
We find that ``diagonal'' (shown by the red circles) performs the best because, on average, they yield the largest $\tau$ and $d$ values with reasonably small standard deviations. The ``minor' method (green circles) performs the second best. The result that the ``diagonal'' works better than ``minor'' is consistent with Theorem~\ref{thm:theorem 2}. The other two types of EWSs perform similarly poorly. Another key observation is that our index, $d$, informs the performance of the EWS in terms of $\tau$ reasonably well and across the four different types of EWSs. 

We show another example in Fig.~\ref{figure 3}(b). In this example, we used the SIS model on a dolphin network.
We gradually increased the infection rate, $D$, as control parameter to induce a bifurcation.
The overall results are similar to those shown in 
Fig.~\ref{figure 3}(a) although the relative performance of the four types of EWSs is not as clear-cut as in the case of Fig.~\ref{figure 3}(a).
We note that, in contrast to Fig.~\ref{figure 3}(a), metric
$d$ is positively correlated with $\tau$ not only across the four types of EWSs but also within each type of EWS.

\begin{figure}[ht]
    \centering
    \includegraphics[width=\linewidth]{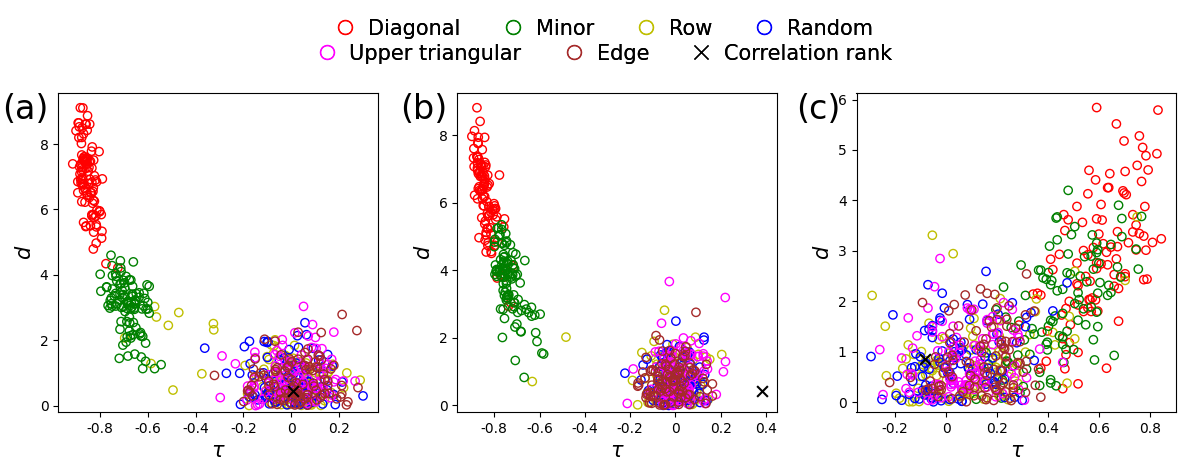}
    \caption{Performance of the ``upper triangular'', ``edge'', and ``correlation rank'' EWSs. We compare these entry selection methods with the four original methods. (a) Coupled double-well dynamics on the lizard network with $60$ nodes, with homogeneous stress and noise, $D$ as the control parameter, and descending simulations. (b) Mutualistic interaction dynamics on the Netsci network with $379$ nodes, with heterogeneous stress and noise, $D$ as the control parameter, and descending simulations. (c) Coupled double-well dynamics on the surfers network with 43 nodes, with heterogeneous stress and noise, with $D$ as the control parameter, and ascending simulations. Note that a small $\tau$ is better in (a) and (b), and a large $\tau$ is better in (c), due to the direction of the simulation (i.e., descending versus ascending). We have arbitrarily chosen these three combinations of dynamics and networks for demonstration.}
    \label{fig:upper-triangle-and-edges}
\end{figure}

The minor matrix contains the diagonal entries. Therefore, we hypothesized that the result that the ``minor'' performs worse than ``diagonal'' but better than ``row'' and ``random'' is mainly due to the fact that ``minor'' partially uses the diagonal entries. To test this hypothesis, we assessed the performance of a different entry selection method called ``upper triangular'', which entails in using only the upper triangular part of the minor matrix excluding the diagonal entries. We showcase this entry selection method in comparison with the others for three combinations of the dynamics and network in Fig.~\ref{fig:upper-triangle-and-edges}. Note that we included in Fig.~\ref{fig:upper-triangle-and-edges} both ascending and descending simulations to test the generality of results against this variation. As expected, the ``upper triangular'' performs as poorly as ``row'' and ``random''.

We also tested the possibility that the sample covariance matrix entries corresponding to edges of the network are more informative than those corresponding to the absence of edges. Specifically, we considered the ``edge'' method to select entries, which entails selecting $n$ edges uniformly at random and use the corresponding entries of $\hat{C}$. Note that this method requires the information on the adjacency matrix, which all the other methods do not. As a substitute of the ``edge'' method when the adjacency matrix is unavailable, we also consider ``correlation rank'', with which we select the top $n$ node pairs in terms of the sample Pearson correlation coefficient (which is readily calculated from $\hat{C}$) at the first value of the control parameter. Then, we use the average of the corresponding $n$ entries of $\hat{C}$ as EWS. The intuition behind this method is that high correlation tends to suggest the existence of edges, although this is not always the case \cite{masuda2023correlation}.

Figure~\ref{fig:upper-triangle-and-edges} shows that the ``edge'' and ``correlation rank'' methods are no better than ``row'', ``random'', and ``upper triangular''. Note that there is only one symbol for ``correlation rank'' because there is just one entry selection using this method; all the other algorithms are stochastic entry selection methods. Given the results shown in Fig.~\ref{fig:upper-triangle-and-edges}, in the following analysis, we focus on the original four entry selection methods with the caveat that a relatively better performance of ``minor'', if we find it, partially owes to the fact that it uses some diagonal entries.

The dominant eigenvalue of the sample covariance matrix has also been employed as EWS \cite{dakos2018identifying, brock2006variance,suweis2014early, chen2019eigenvalues} and depends on both diagonal and off-diagonal entries of the matrix. Therefore, we also compared the dominant eigenvalue method with our proposed EWSs. We have found that the dominant eigenvalue method performs worse than the ``diagonal'' method, further strengthening our claim (see section S9).

We ran a similar analysis for each combination of one of the 23 networks, one of the four dynamical systems models, a control parameter (i.e., either $u$ or $D$), the direction of simulation (i.e., either ascending or descending), whether the stress and noise parameters are both homogeneous or both heterogeneous across the nodes, and one of the four node set selection methods. 
See section S8 for the 23 networks. Then, separately for each dynamical system model, we constructed an ANOVA model aiming to explain $\tau$ or $d$ as a function of these independent variables. 
The goal of this and the subsequent statistical analyses is to show whether the different node set selection methods yield different performances of the EWS. In particular, motivated by the results for the four-node networks and numerical simulations for larger networks, we are interested in showing whether or not ``diagonal'' works the best.

We show the ANOVA results in section S10. The ANOVA model predicts the $\tau$ and $d$ values for the coupled double-well ($R^2 = 0.779$ and $R^2 = 0.618$ for $\tau$ and $d$, respectively), mutualistic interaction ($R^2 = 0.782$ and $R^2 = 0.643$, respectively), and SIS ($R^2 = 0.767$ and $R^2 = 0.656$, respectively) well. However, the model does not fit well for the gene regulatory dynamics ($R^2 = 0.151$ and $R^2 = 0.034$, respectively). This result is consistent with our previous results proposing the ``diagonal'' method \cite{masuda2024anticipating}, where our algorithm works relatively poorly for the gene regulatory dynamics model. The reason for this phenomenon is unclear.  All the independent variables in each ANOVA are highly significant in part due to large sample sizes.

The coefficient on each independent variable obtained by the ANOVA represents how much $\tau$ or $d$ changes in value on average if we change the categorical variable. The results shown in section S10 indicates that, putting aside the gene regulatory dynamics for which the ANOVA fits poorly, the coefficients for the entry selection methods (``minor'', ``row'', or ``random'' relative to ``diagonal'') are larger than the coefficients for other independent variables in most cases. This result implies that the method of entry selection influences more strongly on the value of $\tau$ and $d$ than the other independent variables do. Furthermore, ``diagonal'' is better in terms of the $\tau$ and $d$ values than the other three entry selection methods in all cases, including for the gene regulatory dynamics.

The ANOVA results shown in section S10 also suggests that ``minor'' performs the second best among the four entry selection methods. However, officially, the ANOVA only tells whether the difference between one entry method and a reference method, which we have taken to be ``diagonal'', is significant. To formally verify the casual observation that the ``minor'' performs the second best, we carried out the Tukey's honestly significant difference test
%
%
for each dependent variable (i.e., $\tau$ or $d$) and each dynamics in order to compare the performance among the four entry selection methods.

In Table \ref{HSD}, we find that, for all the four dynamics models, the ``diagonal'' method yields significantly larger $\tau$ and $d$ than the other three entry selection methods. Quantitatively, the difference values (i.e., ``Diff.'' columns in the table) represent the change in $\tau$ or $d$ when one switches from one entry selection method to another. We confirm that ``minor'' is the second best in terms of both $\tau$ and $d$. The ``row'' and ``random'' methods are similarly poorest.

\begin{longtable}{|c|rcc|ccc|}
      \caption{Dependence of $\tau$ and $d$ on the entry selection method for each of the four dynamics. Diff.: Difference between the two entry selection methods. CI: confidence interval.
} 
    \label{HSD}
  \endfirsthead
\endhead
\endfoot
  \hline
  \multirow{2}{*}{} & \multicolumn{3}{|c|}{$\tau$} & \multicolumn{3}{c|}{$d$} \\
  \cline{2-7}
  & \multicolumn{1}{|c}{Diff.} & CI & \multicolumn{1}{c|}{$p$} & Diff. & CI & \multicolumn{1}{c|}{$p$} \\
  \hline
  \multirow{2}{*}{Double-well} & & & & & & \\
    \vspace{-2mm}
  & & & & & & \multicolumn{1}{c|}{}\\
  minor $-$ diag & $-0.154$ & $[-0.159, -0.150]$ & $< 10^{-7}$  & $-2.246$ & $[-2.286, -2.204]$ & $< 10^{-7}$ \\
  row $-$ diag & $-0.666$ & $[-0.671, -0.661]$ & $< 10^{-7}$ & $-3.898$ & $[-3.939, -3.858]$ & $<10^{-7}$  \\
  random $-$ diag & $-0.648$ & $[-0.653, -0.643]$ & $ < 10^{-7}$ & $-3.838$ & $[-3.879, -3.797]$ & $<10^{-7}$ \\
  row $-$ minor & $-0.512$ & $[-0.517, -0.507]$ & $< 10^{-7}$ & $-1.653$ & $[-1.694, -1.612]$ & $< 10^{-7}$ \\
  random $-$ minor & $-0.494$ & $[-0.499, -0.489]$ & $<10^{-7}$ & $-1.592$ & $[-1.633, -1.551]$ & $< 10^{-7}$ \\
  random $-$ row & $0.018$ & $[0.013, 0.023]$ & $< 10^{-7}$ & $0.061$ & $[0.020, 0.102]$ & $0.001$ \\
  \multirow{2}{*}{Mutualistic} & & & & & & \multicolumn{1}{c|}{} \\
  \vspace{-2mm}
  & & & & & & \multicolumn{1}{c|}{}\\
  minor $-$ diag & $-0.133$ & $[-0.141, -0.126]$ & $<10^{-7}$ & $-2.617$ & $[-2.690, -2.543]$ & $<10^{-7}$ \\ 
  row $-$ diag & $-0.692$ & $[-0.700, -0.685]$ & $< 10^{-7}$ & $-4.788$ & $[-4.862, -4.715]$ & $< 10^{-7}$ \\
  random $-$ diag & $-0.669$ & $[-0.677, -0.662]$ & $< 10^{-7}$ & $-4.716$ & $[-4.790, -4.643]$ & $< 10^{-7}$ \\
  row $-$ minor & $-0.559$ & $[-0.566, -0.551]$ & $< 10^{-7}$ & $-2.171$ & $[-2.245, -2.098]$ & $< 10^{-7}$ \\
  random $-$ minor & $-0.536$ & $[-0.544, -0.529]$ & $< 10^{-7}$ & $-2.099$ & $[-2.173, -2.026]$ & $< 10^{-7}$ \\
  random $-$ row & $0.023$ & $[0.015, 0.030]$ & $< 10^{-7}$ & $0.072$ & $[-0.002, 0.145]$ & $0.058$ \\
    \multirow{2}{*}{Gene regulatory} & & & & & & \multicolumn{1}{c|}{} \\
    \vspace{-2mm}
    & & & & & & \multicolumn{1}{c|}{}\\
    minor $-$ diag & $-0.034$ & $[-0.039, -0.028]$ & $< 10^{-7}$ & $-0.094$ & $[-0.126, -0.063]$ & $< 10^{-7}$ \\
    row $-$ diag & $-0.089$ & $[-0.094, -0.084]$ & $< 10^{-7}$ & $-0.174$ & $[-0.205, -0.142]$ & $< 10^{-7}$ \\
    random $-$ diag & $-0.087$ & $[-0.092, -0.082]$ & $< 10^{-7}$ & $-0.172$ & $[-0.203, -0.140]$ & $< 10^{-7}$ \\
    row $-$ minor & $-0.056$ & $[-0.061, -0.050]$ & $< 10^{-7}$ & $-0.077$ & $[-0.185, -0.045]$ & $< 10^{-7}$ \\
    random $-$ minor & $-0.054$ & $[-0.059, -0.048]$ & $< 10^{-7}$ & $-0.078$ & $[-0.109, -0.046]$& $< 10^{-7}$ \\
    random $-$ row & $0.002$ & $[-0.003, 0.007]$ & $0.803$ & $-0.001$ & $[-0.032, 0.031]$ & $1.000$ \\
    \multirow{2}{*}{SIS} & & & & & & \multicolumn{1}{c|}{} \\
    \vspace{-2mm}
    & & & & & & \multicolumn{1}{c|}{}\\
    minor $-$ diag & $-0.124$ & $[-0.131, -0.118]$ & $< 10^{-7}$ & $-2.092$ & $[-2.160, -2.024]$ & $< 10^{-7}$ \\
    row $-$ diag & $-0.555$ & $[-0.561, -0.548]$ & $< 10^{-7}$ & $-3.954$ & $[-4.022, -3.886]$ & $< 10^{-7}$ \\
    random $-$ diag & $-0.542$ & $[-0.548, -0.535]$ & $< 10^{-7}$ & $-3.840$ & $[-3.908, -3.772]$ & $< 10^{-7}$ \\
    row $-$ minor & $-0.430$ & $[-0.437, -0.424]$ & $< 10^{-7}$ & $-1.862$ & $[-1.930, -1.794]$ & $< 10^{-7}$ \\
    random $-$ minor & $-0.417$ & $[-0.424, -0.410]$ & $< 10^{-7}$ & $-1.748$ & $[-1.816, -1.680]$ & $< 10^{-7}$ \\
    random $-$ row & $0.013$ & $[0.006, 0.020]$ & $< 10^{-7}$ & $0.114$ & $[0.046, 0.182]$ & $10^{-4}$ \\    
  \hline  
\end{longtable}

\section{Discussion}
We provided a stochastic differential equation theory to calculate the expectation and fluctuation of the EWS derived from sample covariances. This theory enabled us to compute an index $d$ and the CV of EWSs. With these tools, we provided evidence to support the idea that the use of the sample covariance does not improve the quality of EWSs. This result lends support to the widespread use of the sample variance for EWSs.

There are various directions for future research. First, we have only considered the dynamics that asymptotically approach an equilibrium; examining oscillatory dynamics, which usually accompany a Hopf bifurcation, is left for future work.

Second, degenerate fingerprinting is a closely related method \cite{held2004detection}. As a bifurcation point is approached, the decay rate of one of the modes of the dynamics tends to zero, and the corresponding eigenvector of the Jacobian is the direction in which the state variance diverges. The leading empirical orthogonal function (EOF) aims to approximate this eigenvector using the covariance matrix. Degenerate fingerprinting identifies the leading EOF (or more generally, the largest EOFs) and creates a low-dimensional EWS by projecting the original high-dimensional dynamical system onto the EOFs. Comparing the performance of EOFs with the EWSs examined in this article may be beneficial.

Third, we numerically and mathematically showed positive evidence to support the use of sample variance over sample covariance in the construction of EWSs. However, we only compared random samples from each type of EWS and did not try to optimize the sentinel node set selections. In fact, the EWSs derived from sample covariances with a carefully selected sentinel node set may outperform EWSs only using sample variances. Optimization of sentinel node sets is still an open question for both variance- and covariance-based EWSs.
 
Fourth, in practice, many samples of $x_i^*$ for each sentinel node $i$ may not be available per environment, or control parameter value, contrary to our assumption. Spatial EWSs aim to use spatial statistics, or precisely speaking, statistics over nodes as opposed to over samples of the same nodes to address this situation. We recently showed that major spatial EWSs considered for regular spatial lattices also perform reasonably well on complex networks \cite{maclaren2025applicability}. By definition, spatial EWSs exploit inter-relations among $x_i^*$ at different nodes, as the off-diagonal entries of the sample covariance matrix do. For example, the Moran's $I$ is a spatial covariance measure quantifying similarity between $x_i^*$ and $x_j^*$ for adjacent node pairs $i$ and $j$. Although the EWSs considered in the present study are not spatial EWSs, our results connote that covariation of $x_i$ and $x_j$ may not be helpful for constructing high-quality spatial EWSs. Further investigations of spatial EWSs warrant future work.

Fifth, recent research has been demonstrating the power of machine learning (ML) techniques to construct EWSs for complex nonlinear dynamics \cite{bury2021deep, liu2024early}. In the case of dynamics on networks (e.g., \cite{liu2024early}), ML-based EWSs should be combining information from different nodes to aim at high anticipation performance. The present results may inform design of such ML-based EWSs, because covariance-based features may be better excluded from an ML architecture. Fusion of existing knowledge on EWSs and ML techniques is an interesting topic for further studies.

Data Accessibility. The code is available in Github \cite{data2025}.

Authors' Contributions. S.Y. carried out the theoretical and numerical analyses. S.Y. and Ne.G.M. provided algorithms and codes. Na.M. conceived of, designed, and supervised the study. S.Y. and Na.M. drafted the manuscript. All authors discussed the results, and read, revised, and approved the manuscript.

Competing Interests. The authors declare that they have no competing interests.

Disclaimer. The views expressed in this article are those of the authors and do not reflect the official policy or position of the Department of the Army, DoD, or the U.S. Government.

\section*{Acknowledgments}
Na.M. acknowledges the support provided through the Japan Science and Technology Agency (JST) Moonshot R\&D (grant no. JPMJMS2021), the National Science Foundation (grant no. 2052720),
the National Institute of General Medical Sciences (grant no. 1R01GM148973-01),
and JSPS KAKENHI (grant nos. JP 23H03414, 24K14840, and 24K03013).

\newpage

\setcounter{figure}{0}
\setcounter{table}{0}
\setcounter{section}{0}
\setcounter{equation}{0}

\renewcommand{\thesection}{S\arabic{section}}
\renewcommand{\thefigure}{S\arabic{figure}}
\renewcommand{\thetable}{S\arabic{table}}
\renewcommand{\theequation}{S\arabic{equation}}
\begin{center}
\vspace*{12pt}
{\LARGE Supplementary Information for: Using covariance of node states to design early warning signals for network dynamics\\
\vspace{12pt}
\Large Shilong Yu, Neil G. MacLaren, and Naoki Masuda}
\vspace{10pt} \\
\end{center}


\renewcommand{\thesection}{S\arabic{section}}
\renewcommand{\thesubsection}{S\arabic{section}.\arabic{subsection}}
\renewcommand{\thetable}{S\arabic{table}}
\renewcommand{\thefigure}{S\arabic{figure}}

\renewcommand{\theequation}{S\arabic{equation}}

\section{Dynamical system models}\label{sec: s9}

In this section, we explain the other three dynamical system models than the coupled double-well model used in our numerical simulations.

\subsection{Mutualistic interaction dynamics}

The stochastic mutualistic interaction dynamics is given by
\begin{equation}
    \text{d}x_i = \left[B_i + x_i\left(1 - \frac{x_i}{K_i}\right)\left(\frac{x_i}{\tilde{C}_i}-1\right) + D\sum_{j = 1}^N A_{ij}\frac{x_ix_j}{\tilde{D}_i + E_i x_i + H_j x_j}\right]\text{d}t + \sigma_i \text{d}W_i,
\end{equation}
where $x_i$ is the population of the $i$th species (with $i \in \{1, \ldots, N\}$), $B_i$ is the migration rate, $K_i$ is the carrying capacity, $\tilde{C}_i$ is the Allee constant, and $\tilde{D}_i$, $E_i$, and $H_j$ affect the strength of the mutual interaction between species $i$ and $j$. Following \cite{gao2016universal}, we set $B_i = 0.1 + u + \Delta u_i$, $\tilde{C}_i = 1$, $\tilde{D}_i = 5$, $E_i = 0.9$, $H_i = 0.1$, and $K_i = 5$, $\forall i \in \{1, \ldots, N\}$. As in \cite{masuda2024anticipating}, when $u$ is the control parameter, we initially set $u = 0$ and $D = 1$, and only decrease $u$; when $D$ is the control parameter, we initially set $u = 0$ and $D = 1$, and only decrease $D$. We set $\sigma = 0.25$ \cite{masuda2024anticipating}.

\subsection{Gene regulatory dynamics}

The stochastic gene regulatory dynamics is given by 
\begin{equation}
    \text{d}x_i = \left(-Bx_i^f + D\sum_{j = 1}^N A_{ij}\frac{x_j^h}{x_j^h + 1} + u_i\right)\text{d}t + \sigma_i \text{d}W_i,
\end{equation}
where $x_i$ is the expression level of the $i$th gene. Following \cite{gao2016universal}, we set $B = 1$, $f = 1$, and $h = 2$. We also set  $\sigma = 5 \times 10^{-6}$, following \cite{masuda2024anticipating}. Similarly to  \cite{masuda2024anticipating}, when $u$ is the control parameter, we initially set $u = 0$ and $D = 1.5$, and only decrease $u$; when $D$ is the control parameter, we initially set $u = 0$ and $D = 1$, and only decrease $D$. The only difference between the values used here and \cite{masuda2024anticipating} is that we changed $D$ from $1$ to $1.5$ for the same reason that we changed $D$ from $0.05$ to $0.02$ for the coupled double-well dynamics.

\subsection{Susceptible-infectious-susceptible (SIS) dynamics}

We use a stochastic SIS dynamics given by
\begin{equation}
    \text{d}x_i = \left[D\sum_{j = 1}^N A_{ij}(1-x_i)x_j - \mu x_i\right]\text{d}t + \sigma_i \text{d}W_i,
\end{equation}
where $x_i$ is the probability that the $i$th node is infectious, $D$ is the rate at which $i$ is infected by an infectious neighbor, and $\mu$ is the rate at which $i$ recovers. We set $\mu = 1$ and only use $D$ as the control parameter. We do not use $u$ as control parameter because it would be unrealistic that infectious individuals spontaneously emerge. We initially set $D$ to $0$ and $0.7$ in the ascending and descending simulations, respectively, and $\sigma = 5 \times 10^{-4}$, following \cite{masuda2024anticipating}.

\newpage

\section{Kendall's $\tau$}\label{sec: s10}

Kendall's $\tau$ is a rank correlation coefficient. To analyze which entry combination $S$ best signals an impending regime shift, we compute $\tau$ between the given EWS and the control parameter, as is done in many studies of EWSs \cite{dakos2010spatial, kefi2014early}. 
For expository purposes, we consider the case in which $u$ gradually increases (i.e., ascending simulations).
For a given $S$, we compute $\tau$ between the EWS at $u = 0$, that at $u = \Delta u$, $\ldots$, that at $u= (\tilde{K}-1)\Delta u$, $\tilde{K}$ is the number of control parameter values used, and the corresponding sequence of the control parameter value, i.e., $0$, $\Delta u$, $\ldots$, $(\tilde{K}-1) \Delta u$. A monotonically increasing EWS yields $\tau = 1$ when the control parameter linearly increases (i.e., ascending simulations) and suggests that the EWS is good.
A good EWS corresponds to $\tau = -1$ when the control parameter linearly decreases, i.e., in descending simulations. If the EWS is not responsive to the gradual change in the control parameter, we would obtain $\tau$ close to $0$. 

\newpage

\section{Simulation methods}\label{sec: s11}

For each simulation, regardless of the control parameter being $u$ or $D$, we set the initial value of $x_i$, $\forall i \in \{1, \ldots, N\}$, to $1$ and $5$ in the ascending and descending simulations, respectively, of the coupled double-well dynamics, $5$ for the descending simulations of the mutualistic interaction and gene regulatory dynamics, and $0.001$ and $0.999$ for the ascending and descending simulations, respectively, of the SIS dynamics~\cite{masuda2024anticipating}. It should be noted that, for mutualistic interaction dynamics, we carry out descending but not ascending simulations because a common interest in ecology is to model species collapse and loss of resilience of ecosystems \cite{maclaren2025applicability, scheffer2009early}. Similarly, for the gene-regulatory dynamics, we carry out descending but not ascending simulations because we are interested in modeling the loss of resilience of the active state \cite{maclaren2025applicability, gao2016universal}.

We used the Euler-Maruyama method with $\Delta t = 0.01$ to simulate each dynamics. For the mutualistic interaction, gene regulatory, and SIS dynamics, if $x_i(t) < 0$ at any $t$ due to dynamical noise, we set $x_i(t) = 0$ because  $x_i(t) < 0$ is unrealistic. Given a value of the control parameter, each simulation lasts till 200 time units (TU), i.e., till $t=200$. The first 100 TUs is considered to bring the system to equilibrium. After that, we took $L = 100$ evenly spaced samples from $x_i(t)$, $\forall i \in \{1, \ldots, N\}$ as $x_{i, \ell}$, $\ell \in \{1, \ldots, L\}$, for calculating sample variance and covariances. The distance between adjacent samples is $1$ TU. 
Using the $L$ sampled points, we form the $N \times N$ unbiased sample covariance matrix, $\hat{C}$, and use it to compute EWSs. In fact, the calculation of each type of EWS only needs an $n\times n$ minor matrix of $\hat{C}$, the calculation of which one only needs $x_{i, \ell}$, $\ell \in \{1, \ldots, L\}$ from $n$ values of $i$ (i.e., $n$ nodes).

Regarding the simulation direction, we linearly increased or decreased the control parameter.
For example, if the control parameter is $u$ and the direction of simulation is ascending, then we run a simulation for each $u \in \{0, 0 + \Delta u, 0 + 2\Delta u, \ldots\}$ in this order. At each $u$ value, we produce $\hat{C}$ and calculate the EWSs.
We followed \cite{masuda2024anticipating} to set $(\Delta u, \Delta D) = (0.025, 0.0025)$
and $(-0.025, -0.0025)$ in the ascending and descending simulations, respectively, of the coupled double-well dynamics,
$(\Delta u, \Delta D) = (-0.1, -0.01)$ in the descending simulations of the mutualistic interaction dynamics,
$(\Delta u, \Delta D) = (-0.01, -0.01)$ in the descending simulations of the gene regulatory dynamics, and
$\Delta D = 0.0025$ and $-0.0025$ in the ascending and descending simulations, respectively, of the SIS dynamics. 

In either ascending or descending simulations, we stop increasing or decreasing the value of the control parameter when
at least one node has transited to a qualitatively different state, i.e., from the lower to upper state in the case of an ascending simulation and the converse in the case of a descending simulation. To determine this, 
we define node $i$ to be no longer near its initial state if $x_i \geq 3$ for the ascending simulations of the coupled double-well dynamics, $x_i \leq 3$ for the descending simulations of the coupled double-well dynamics, $x_i < 0.1$ for the descending simulations of the mutualistic interaction dynamics, $x_i < 0.1$ for the descending simulations of the gene regulatory dynamics, $x_i \geq 0.1$ for the ascending simulations of the SIS dynamics, and $x_i \leq 0.1$ for the descending simulations of the SIS dynamics.   

We calculate $d$ as follows \cite{masuda2024anticipating}. Suppose as an example that we simulate the dynamics at $u = \overline{u}_k:= 0 + k\Delta u$, $k \in \{0, 1, \cdots, \tilde{K}-1\}$, where $\overline{u}_{\tilde{K}-1}$ is the last control parameter value before at least one node is no longer near its initial state. We then define $k^{(1)}:= \text{round} (0.1\tilde{K})$ and $k^{(2)}:= \text{round}(0.9\tilde{K})$ so that $k^{(1)}$ is relatively far from the first regime shift and $k^{(2)}$ is relatively close to the same regime shift. Then, we calculate $\hat{C}^{(1)}$ and $\hat{C}^{(2)}$ from the $L$ samples obtained from the simulations at the $k^{(1)}$th and $k^{(2)}$th control parameter values, respectively. When $D$ is the control parameter, we similarly produce $\hat{C}^{(1)}$ and $\hat{C}^{(2)}$. Then, we calculate the mean and standard deviation of the EWS at $u = \overline{u}_{k^{(1)}}$ and $u = \overline{u}_{k^{(2)}}$ using Eqs. (3.7) and  (3.8) in the main text. Finally, we obtain $d$ using Eq.~(2.2) in the main text.

\newpage

\section{Derivation of $\mathbb{E}\left[\hat{C}_{i,j}\right]$ and $\text{var}\left[{\hat{C}_{i,j}}\right]$}\label{sec: Appendix A}	

We derive Eq.~(3.5) in the main text using similar techniques to the ones in \cite{masuda2024anticipating}. Define $\bar{z_i} = \sum_{\ell=1}^L z_{i, \ell} / L$. Then, we obtain
	\begin{equation}
		\begin{aligned}
			\mathbb{E}[\hat{C}_{i,j}] =& \frac{1}{L-1}\mathbb{E}\left[\sum_{\ell = 1}^L (z_{i, \ell} - \bar{z_i})(z_{j, \ell} - \bar{z_j})\right]
			\\ =& \frac{1}{L-1}\mathbb{E}\left[\sum_{\ell = 1}^L \left(z_{i, \ell}z_{j,\ell} - \bar{z_i}z_{j,\ell} - \bar{z_j}z_{i,\ell} + \bar{z_i}\bar{z_j}\right)\right] \\ =& \frac{1}{L-1}\mathbb{E}\left[\sum_{\ell = 1}^L z_{i, \ell}z_{j,\ell} - \frac{1}{L}\sum_{\ell = 1}^L (z_{i,1} + \cdots + z_{i, L})z_{j,\ell} - \frac{1}{L}\sum_{\ell = 1}^L (z_{j,1} + \cdots + z_{j,L})z_{i,\ell} \right. \\ &+ \left.
			\frac{1}{L^2}\sum_{\ell = 1}^L (z_{i, 1} + \cdots + z_{i, L})(z_{j, 1} + \cdots + z_{j, L})\right] \\ 
			=& \frac{1}{L-1}\mathbb{E}\left[\sum_{l = 1}^L z_{i, \ell}z_{j, \ell} - \frac{2}{L}\sum_{\ell = 1}^L z_{i, \ell}z_{j, \ell} + \frac{1}{L}\sum_{\ell = 1}^L z_{i, \ell}z_{j, \ell}\right]
			\\ =& \frac{L-1}{(L-1)L}\mathbb{E}\left[\sum_{\ell = 1}^L z_{i,\ell}z_{j,\ell}\right] \\
			 =& \frac{1}{L}\mathbb{E}\left[\sum_{\ell = 1}^L z_{i,\ell}z_{j,\ell}\right] \\
			 =& C_{ij}.
		\end{aligned}
	\label{eq:E[hatC]}	
	\end{equation}
	
We now derive Eq.~(3.6) in the main text. Because
	\begin{equation}
		(\hat{C}_{i,j})^2 = \frac{1}{(L-1)^2}\left[\sum_{\ell = 1}^L z_{i,\ell}z_{j,\ell} - \frac{1}{L}(z_{i,1} + \cdots + z_{i,L})(z_{j,1} + \cdots + z_{j,L})\right]^2,
	\end{equation}
we obtain	
	\begin{equation}\label{ECov_ij^2}
		\begin{aligned}
			\mathbb{E}[(\hat{C}_{i,j})^2] =& \frac{1}{(L-1)^2}\mathbb{E}\left[\left(\sum_{\ell = 1}^L z_{i,\ell}z_{j,\ell}\right)^2\right] \\ &+ \frac{1}{(L-1)^2} \mathbb{E}\left[2\left(\sum_{l = 1}^L z_{i,\ell}z_{j,\ell}\right)\left(-\frac{1}{L}\right)(z_{i,1} + \cdots + z_{i,L})(z_{j,1} + \cdots + z_{j,L}) \right]  \\& + \frac{1}{(L-1)^2}\mathbb{E}\left[\frac{1}{L^2}(z_{i,1} + \cdots + z_{i,L})^2(z_{j,1} + \cdots + z_{j,L})^2\right].
		\end{aligned}
	\end{equation}
We calculate the first term on the right-hand side of Eq.~\eqref{ECov_ij^2} as follows:
	\begin{equation}
		\mathbb{E}\left[\left(\sum_{\ell = 1}^L z_{i,\ell}z_{j,\ell}\right)^2\right] = L\mathbb{E}[z_i^2 z_j^2] + L(L-1)\mathbb{E}[z_iz_j]^2,
	\end{equation}
where we used $\mathbb{E}[z_{i,\ell}^2z_{j,\ell}^2] = \mathbb{E}[z_i^2z_j^2]$ and $\mathbb{E}[z_{i,\ell}z_{j,\ell}z_{i,\ell'}z_{j,\ell'}] = \mathbb{E}[z_{i}z_{j}]^2$ for any $i, j \in \{1, \ldots, N\}$ and $\ell, \ell' \in \{1, \ldots, L\}$ with $\ell \neq \ell'$. 	
We calculate the second term on the right-hand side of Eq.~\eqref{ECov_ij^2} as follows:
 	\begin{equation}
 		\begin{aligned}
 			&\mathbb{E}\left[\left(\sum_{\ell = 1}^L z_{i,\ell}z_{j,\ell}\right)(z_{i,1} + \cdots + z_{i,L})(z_{j,1} + \cdots + z_{j,L})\right] \\ &= \sum_{\ell= 1}^L \mathbb{E}\left[z_{i,\ell}z_{j,\ell}z_{i,\ell}z_{j,\ell}\right] + \sum_{\ell = 1}^L \mathbb{E}\left[z_{i,\ell}z_{j,\ell}\sum_{\ell'=1,\ell'\neq \ell}^L z_{i,\ell'}z_{j,\ell'}\right]
 			\\ &= L\mathbb{E}[z_i^2z_j^2] + \sum_{\ell = 1}^L (L-1)\mathbb{E}[z_{i,\ell}z_{j,\ell}]\mathbb{E}[z_{i,\ell'}z_{j,\ell'}] \\ &= L\mathbb{E}[z_i^2z_j^2] + L(L-1)\mathbb{E}[z_iz_j]^2.
 		\end{aligned}
 	\end{equation} 	
The third term simplifies as follows:
 	\begin{equation}
 		\begin{aligned}
 			&\mathbb{E}\left[(z_{i,1} + \cdots + z_{i,L})^2(z_{j,1} + \cdots + z_{j,L})^2\right] \\ &= L\mathbb{E}[z_i^2z_j^2] + \frac{L(L-1)}{2}\cdot 2\mathbb{E}[z_i^2]\mathbb{E}[z_j^2] + \frac{L(L-1)}{2} \cdot 4\mathbb{E}[z_iz_j]^2
 		\end{aligned}
 	\end{equation}
 	because $\mathbb{E}[z_{i,\ell}^2z_{j,\ell'}^2] = \mathbb{E}[z_i^2]\mathbb{E}[z_j^2]$ for $\ell \neq \ell'$. We know $\mathbb{E}[z_i^2] = C_{ii}$, $\mathbb{E}[z_iz_j] = C_{ij}$, $\mathbb{E}[z_i^2z_j^2] = C_{ii}C_{jj} + 2(C_{ij})^2$ \cite{masuda2024anticipating, vatiwutipong2019alternative}. Using these relationships, we obtain
 	\begin{equation}\label{covijcovij}
 		\begin{aligned}
 			\mathbb{E}[(\hat{C}_{i,j})^2] =& \frac{1}{(L-1)^2} \left\{L\left[C_{ii}C_{jj} + 2(C_{ij})^2 \right] + L(L-1)(C_{ij})^2 \right.\\ &  - \frac{2}{L}\left. \left\{L\left[C_{ii}C_{jj} + 2(C_{ij})^2\right] + L(L-1)(C_{ij})^2\right\}\right\}
 		\\ &+ \frac{1}{(L-1)^2}\frac{1}{L^2}\left\{L\left[ C_{ii}C_{jj}+2(C_{ij})^2 \right] + L(L-1)C_{ii}C_{jj} + 2L(L-1)(C_{ij})^2\right\}
 		\\ =& \frac{1}{(L-1)^2}\left\{ \left( L - 2 + \frac{1}{L} + \frac{L-1}{L}\right) C_{ii}C_{jj} \right. \\ & \left. + \left[ 2L + L(L-1) - 4 - 2(L-1) + \frac{2}{L} + \frac{2L-2}{L}\right] (C_{ij})^2\right\}
 		\\ =& \frac{1}{L-1}\left[ C_{ii}C_{jj} + L(C_{ij})^2 \right].
 		\end{aligned}
 	\end{equation}
Using Eqs.~\eqref{eq:E[hatC]} and \eqref{covijcovij}, we obtain
 	\begin{equation}
 		\begin{aligned}
 			\text{var}[\hat{C}_{i,j}] &= \mathbb{E}\left[ \left(\hat{C}_{i,j}\right)^2 \right] - \left( \mathbb{E}[\hat{C}_{i,j}] \right)^2 \\
			&= \frac{1}{L-1}(C_{ii}C_{jj} + LC_{ij}^2) - (C_{ij})^2 \\ &= \frac{1}{L-1}\left[ C_{ii}C_{jj} + (C_{ij})^2\right].
 		\end{aligned}
 	\end{equation}

\newpage

\section{Derivation of $\text{var}\left(\frac{1}{n}\sum_{m = 1}^n\hat{C}_{i_m, j_m} \right)$}

In this section, we derive Eq.~(3.8) in the main text. We first compute $\mathbb{E}[\hat{C}_{i,j}\hat{C}_{i',j'}]$ as follows:
\begin{equation}\label{ECovijCovi'j'}
	\begin{aligned}   (L-1)^2\mathbb{E}\left[\hat{C}_{i,j}\hat{C}_{i',j'}\right] =& \mathbb{E}\left[ \left(\sum_{\ell = 1}^L z_{i,\ell}z_{j,\ell} - \frac{1}{L}(z_{i,1} + \cdots z_{i,L})(z_{j,1} + \cdots + z_{j,L})\right) \times \right. \\
	 & \left. \left(\sum_{\ell = 1}^L z_{i',l}z_{j',l}-\frac{1}{L}(z_{i',1} + \cdots + z_{i',L})(z_{j',1} + \cdots + z_{j',L})\right)\right] \\
=& \mathbb{E}\left[\left(\sum_{\ell = 1}^L z_{i,\ell}z_{j,\ell}\right)\left(\sum_{\ell = 1}^L z_{i',\ell}z_{j',\ell}\right)\right] \\
& - \mathbb{E}\left[\left(\frac{1}{L}\sum_{\ell = 1}^L z_{i,\ell}z_{j,\ell}(z_{i',1} + \cdots + z_{i',L})(z_{j'1} + \cdots + z_{j',L})\right)\right] \\
& - \mathbb{E}\left[\frac{1}{L}\sum_{\ell = 1}^L z_{i',\ell}z_{j',\ell}(z_{i,1} + \cdots + z_{i,L})(z_{j,1} + \cdots + z_{j,L})\right] \\
&+ \frac{1}{L^2}\mathbb{E}\left[ (z_{i,1} + \cdots + z_{i,L})(z_{j,1} + \cdots + z_{j,L}) \times \right. \\
& \left. (z_{i',1} + \cdots + z_{i',L})(z_{j',1} + \cdots + z_{j',L}) \right].
	\end{aligned}
\end{equation}
We calculate the first term on the right-hand side (i.e., after the second equality) of Eq.~\eqref{ECovijCovi'j'} as follows:
 	\begin{equation}\label{49}
 		\begin{aligned}
 			\mathbb{E}\left[\left(\sum_{\ell = 1}^L z_{i,\ell}z_{j,\ell}\right)\left(\sum_{\ell = 1}^L z_{i',\ell}z_{j',\ell}\right)\right] &= \mathbb{E}\left[\sum_{\ell = 1}^L z_{i,\ell}z_{j,\ell}z_{i',\ell}z_{j',\ell}\right] + \mathbb{E}\left[\sum_{\ell = 1, \ell'=1, \ell \neq \ell'}^Lz_{i,\ell}z_{j,\ell}z_{i',\ell'}z_{j',\ell'}\right] 
 			\\ &= L\mathbb{E}[z_iz_jz_{i'}z_{j'}] + L(L-1)\mathbb{E}[z_iz_j]\mathbb{E}[z_{i'}z_{j'}].
 		\end{aligned}
 	\end{equation}
We obtain the second term on the same right-hand side as follows:
 	\begin{equation}\label{50}
 		\begin{aligned}
& \mathbb{E}\left[\left(\frac{1}{L}\sum_{\ell = 1}^L z_{i,\ell}z_{j,\ell}(z_{i',1} + \cdots + z_{i',L})(z_{j'1} + \cdots + z_{j',L})\right)\right] \\
=& \frac{1}{L}\mathbb{E}\left[\sum_{\ell = 1}^L z_{i,\ell}z_{j,\ell}z_{i',\ell}z_{j',\ell} + \sum_{\ell = 1}^L z_{i,\ell}z_{j,\ell}\sum_{\ell'=1,\ell'\neq l}^L z_{i',\ell'}z_{j',\ell'}\right] \\
=& \frac{1}{L} \cdot L \mathbb{E}[z_iz_jz_{i'}z_{j'}] + \frac{1}{L} \cdot L\mathbb{E}[z_iz_j](L-1)\mathbb{E}[z_{i'}z_{j'}].
 		\end{aligned}
 	\end{equation}
Similarly, the third term is simplified as follows:
 	\begin{equation}\label{51}
  \begin{aligned}
& \mathbb{E}\left[\frac{1}{L}\sum_{\ell = 1}^L z_{i',\ell}z_{j',\ell}(z_{i,1} + \cdots + z_{i,L})(z_{j,1} + \cdots + z_{j,L})\right] \\
=& \frac{1}{L}\mathbb{E}\left[\sum_{\ell = 1}^L z_{i',\ell}z_{j',\ell}z_{i,\ell}z_{j,\ell} + \sum_{\ell = 1}^L z_{i',\ell}z_{j',\ell}\sum_{\ell' = 1, \ell' \neq \ell}^L z_{i,\ell'}z_{j,\ell'}\right]\\
=& \frac{1}{L} \cdot L \mathbb{E}[z_iz_jz_{i'}z_{j'}] + \frac{1}{L} \cdot L\mathbb{E}[z_iz_j](L-1)\mathbb{E}[z_{i'}z_{j'}].
   \end{aligned}
 	\end{equation}
We calculate the fourth term as follows:
 	\begin{equation}\label{52}
 		\begin{aligned}
& \frac{1}{L^2}\mathbb{E}\left[ (z_{i,1} + \cdots + z_{i,L})(z_{j,1} + \cdots + z_{j,L}) 
(z_{i',1} + \cdots + z_{i',L})(z_{j',1} + \cdots + z_{j',L}) \right] \\
=& \frac{1}{L^2}\mathbb{E}\left[\sum_{\ell = 1}^L z_{i,\ell}z_{j,\ell}z_{i',\ell}z_{j',\ell} + \sum_{\ell = 1, \ell'=1, \ell \neq \ell'}^Lz_{i,\ell}z_{j,\ell}z_{i',\ell'}z_{j',\ell'} + \cdots \right]
\\ =& \frac{1}{L^2}\cdot L \mathbb{E}[z_iz_jz_{i'}z_{j'}] + \frac{1}{L^2} \cdot L(L-1)(\mathbb{E}[z_iz_j]\mathbb{E}[z_{i'}z_{j'}] + \mathbb{E}[z_iz_{i'}]\mathbb{E}[z_{j}z_{j'}] + \mathbb{E}[z_iz_{j'}]\mathbb{E}[z_{i'}z_{j}]).
 		\end{aligned}
 	\end{equation}
 	By Isserlis' theorem and by the fact that the stationary distribution of a multivariate OU process is multivariable Gaussian \cite{isserlis1918formula, vatiwutipong2019alternative}, it holds true that
\begin{equation}
\mathbb{E}[z_iz_jz_{i'}z_{j'}] = C_{ij}C_{i'j'} + C_{ii'}C_{jj'} + C_{ij'}C_{i'j}.
\label{eq:Isseris}
\end{equation}
By applying Eq.~\eqref{eq:Isseris} to Eqs.~\eqref{49}, \eqref{50}, \eqref{51}, and \eqref{52} and substituting the results in Eq.~\eqref{ECovijCovi'j'}, we obtain
 	\begin{equation}\label{54}
 		\begin{aligned}
 			(L-1)^2\mathbb{E}[\hat{C}_{i,j}\hat{C}_{i',j'}] =& L(C_{ij}C_{i'j'} + C_{ii'}C_{jj'} + C_{ij'}C_{ji'}) + L(L-1)C_{ij}C_{i'j'} \\ &-2(C_{ij}C_{i'j'} + C_{ii'}C_{jj'} + C_{ij'}C_{ji'}) - 2(L-1)C_{ij}C_{i'j'} \\ &+ \frac{1}{L}(C_{ij}C_{i'j'} + C_{ii'}C_{jj'} + C_{ij'}C_{ji'}) \\ &+ \frac{L-1}{L}(C_{ij}C_{i'j'} + C_{ii'}C_{jj'} + C_{ij'}C_{i'j}) \\=& (L-1)^2C_{ij}C_{i'j'} + (L-1)C_{ii'}C_{jj'} + (L-1)C_{ij'}C_{ji'},
 		\end{aligned}
 	\end{equation}
which leads to
	\begin{equation}\label{covijcovi'j'}
		\mathbb{E}[\hat{C}_{i,j}\hat{C}_{i',j'}] = \frac{1}{L-1}\left[ (L-1)C_{ij}C_{i'j'} + C_{ii'}C_{jj'} + C_{ij'}C_{i'j}\right].
	\end{equation}

Now we are ready to derive Eq.~(3.8) in the main text. We first observe that 
\begin{equation}\label{ECimjmsquared}
    \mathbb{E}\left[\left(\frac{1}{n}\sum_{m = 1}^n \hat{C}_{i_m, j_m}\right)^2\right] = \frac{1}{n^2}\mathbb{E}\left[\sum_{m = 1}^n \left(\hat{C}_{i_m, j_m}\right)^2 + \sum_{m, m'=1, m \neq m'}^n \hat{C}_{i_m, j_m}\hat{C}_{i_{m'}, j_{m'}}\right].
\end{equation}
By substituting Eqs.~\eqref{covijcovij} and \eqref{covijcovi'j'} in Eq.~\eqref{ECimjmsquared}, we obtain
	\begin{equation}\label{average Covimjm}
		\begin{aligned}
			&\mathbb{E}\left[\left(\frac{1}{n}\sum_{m = 1}^n \hat{C}_{i_m,j_m}\right)^2\right] \\
			=& \frac{1}{n^2}\left\{\sum_{m = 1}^n \frac{1}{L-1}\left[C_{i_mi_m}C_{j_mj_m} + L(C_{i_mj_m})^2\right]\right\} \\
			& + \frac{1}{n^2}\left\{ \sum_{m, m'=1, m \neq m'}^n\frac{1}{L-1}\left[ (L-1)C_{i_mj_m}C_{i_{m'}j_{m'}} + C_{i_mi_{m'}}C_{j_mj_{m'}} + C_{i_mj_{m'}}C_{i_{m'}j_m} \right] \right\}
		\end{aligned}
\end{equation}
and
\begin{equation}
	\begin{aligned}\label{1/k covij squared}
			\mathbb{E}\left[\frac{1}{n}\sum_{m = 1}^n \hat{C}_{i_m,j_m}\right]^2 &= \frac{1}{n^2}\left(\sum_{m = 1}^n C_{i_mj_m}\right)^2  \\ &= \frac{1}{n^2}\sum_{m = 1}^n \left(C_{i_mj_m}\right)^2  +\frac{1}{n^2}\sum_{m, m'=1, m \neq m'}^n C_{i_mj_m}C_{i_{m'}j_{m'}}.
	\end{aligned}
  \end{equation}
Using Eqs.~\eqref{average Covimjm} and \eqref{1/k covij squared}, we obtain
 \begin{equation}\label{last equation}
	\begin{aligned}
	&\text{var}\left[\frac{1}{n}\sum_{m = 1}^n \hat{C}_{i_m, j_m}\right] \\
	=& \mathbb{E}\left[\left(\frac{1}{n}\sum_{m = 1}^n \hat{C}_{i_m, j_m}\right)^2\right] - \mathbb{E}\left[\sum_{m = 1}^n \hat{C}_{i_m, j_m}\right]^2 \\
	 =& \frac{1}{n^2(L-1)}\left\{\sum_{m = 1}^n\left[C_{i_mi_m}C_{j_mj_m} + (C_{i_mj_m})^2\right] + \sum_{m,m'=1, m \neq m'}^n \left(C_{i_mi_{m'}}C_{j_mj_{m'}} + C_{i_mj_{m'}}C_{i_{m'}j_m}\right)\right\} \\ =& \frac{1}{n^2(L-1)}\left[\sum_{m,m'=1}^n C_{i_mi_{m'}}C_{j_mj_{m'}} + \sum_{m,m'=1}^n C_{i_mj_{m'}}C_{i_{m'}j_m}\right].
	\end{aligned}
\end{equation}

\newpage

\section{Derivation of entries in the covariance matrix for networks with four nodes}\label{sec: Appendix C}

\subsection{Chain}\label{c1chain}

The Lyapunov equation reads
\begin{equation}
\begin{aligned}
    &\begin{pmatrix}
		-2 x_1^* & -w & 0 & 0 \\
		-w & -2 x_2^* & -w & 0 \\
		0 & -w & -2 x_2^* & -w \\
		0 & 0 & -w & -2 x_1^*
    \end{pmatrix}\begin{pmatrix}
        C_{11} & C_{12} & C_{13} & C_{14} \\
        C_{21} & C_{22} & C_{23} & C_{24} \\
        C_{31} & C_{32} & C_{33} & C_{34} \\
        C_{41} & C_{42} & C_{43} & C_{44}
    \end{pmatrix} \\ &+ \begin{pmatrix}
        C_{11} & C_{12} & C_{13} & C_{14} \\
        C_{21} & C_{22} & C_{23} & C_{24} \\
        C_{31} & C_{32} & C_{33} & C_{34} \\
        C_{41} & C_{42} & C_{43} & C_{44}
    \end{pmatrix} \begin{pmatrix}
		-2 x_1^* & -w & 0 & 0 \\
		-w & -2 x_2^* & -w & 0 \\
		0 & -w & -2 x_2^* & -w \\
		0 & 0 & -w & -2 x_1^*
    \end{pmatrix} = \begin{pmatrix}
        \sigma_1^2 & 0 & 0 & 0 \\
        0 & \sigma_2^2 & 0 & 0 \\
        0 & 0 & \sigma_2^2 & 0 \\
        0 & 0 & 0 & \sigma_1^2
    \end{pmatrix},
\end{aligned}
\end{equation}
which gives us a system of $16$ linear equations and $16$ unknowns $C_{ij}$, $i, j \in \{1, 2, 3, 4\}$. Because $C$ is a symmetric matrix, we can choose to consider the equations only involving the diagonal and the upper triangular entries of $C$, thereby reducing the total number of unknowns to $10$. By symmetry of the network, we obtain $C_{11} = C_{44}$, $C_{22} = C_{33}$, $C_{12} = C_{34}$, and $C_{13} = C_{24}$. Therefore, we only need to solve the following set of six linear equations for $C_{11}$, $C_{12}$, $C_{13}$, $C_{14}$, $C_{22}$, and $C_{23}$:
\begin{align}
    -2x_1^*C_{11} - wC_{12} - 2x_1^*C_{11} - wC_{12} &= \sigma_1^2, \\
    -2x_1^*C_{12} - wC_{22} - wC_{11} - 2x_2^*C_{12} - wC_{13} &= 0, \\
    -2x_1^*C_{13} - wC_{23} - wC_{12} - 2x_2^*C_{13} - wC_{14} &= 0, \\
    -2x_1^*C_{14} - wC_{13} - wC_{13} - 2x_1^*C_{14} &= 0, \\
    -wC_{12} - 2x_2^*C_{22} - wC_{32} - wC_{21} - 2x_2^*C_{22} - wC_{23} &= \sigma_2^2, \\
    -wC_{13} - 2x_2^*C_{23} - wC_{22} - wC_{22} - 2x_2^*C_{23} - wC_{13} &= 0. 
\end{align}

We use Python to solve this system. The explicit solution is long, so we do not show it. For the same reason, we do not show the explicit solution of each entry of $C$ for the star and lollipop graphs in the following sections.

\subsection{Star}\label{c2star}

The system of stochastic differential equations is given by
 \begin{align}
		dx_1 &= [f(x_1) + w(x_2 + 1) + w(x_3+1) + w(x_4+1)]dt + \sigma_1 dW_1, \\
		dx_2 &= [f(x_2) + w(x_1 + 1)]dt + \sigma_2 dW_2, \\
		dx_3 &= [f(x_3) + w(x_1 + 1)]dt + \sigma_3 dW_3, \\
		dx_4 &= [f(x_4) + w(x_1 + 1)]dt + \sigma_4 dW_4.
 \end{align}
The equilibrium in the absence of noise satisfies
 \begin{align}	
		(x_1^*)^2 + r + w(x_2^* + 1) + w(x_3^*+1) + w(x_4^*+1) &= 0, \label{1st star graph}\\ 
		(x_2^*)^2 + r + w(x_1^* + 1) &= 0, \label{2nd star graph}\\
		(x_3^*)^2 + r + w(x_1^* + 1) &= 0, \\
		(x_4^*)^2 + r + w(x_1^* + 1) &= 0. 
 \end{align}
Due to symmetry in the network structure and $-1 \le x_i^* < 0$ for the stable equilibrium, we obtain $x_2^* = x_3^* = x_4^*$. Using $x_2^* = x_3^* = x_4^*$ $(<0)$, we reduce Eq.~\eqref{1st star graph} to
\begin{equation}
	(x_1^*)^2 + r + 3w(x_2^* + 1) = 0.
	\label{eq:1st-star-graph-sym}
\end{equation}
As we gradually increase $r$ from $-1$, the number of intersections of the two parabolas represented by Eqs.~\eqref{2nd star graph} and \eqref{eq:1st-star-graph-sym} changes from four to two, and then to zero. Define $r_c'$ and $r_c$ as the value of $r$ at which the number of intersections changes from four to two and from two to zero, respectively. We obtain $r_c' \approx -0.210$ and $r_c \approx -0.112$.
We calculate the $d$ values at $r =-0.22$ and $r = -0.4$ so that one $r$ value is close to $r_c'$ and the other is far from $r_c'$.

 The sign-flipped Jacobian is \begin{equation}
		A = \begin{pmatrix}
			-2 x_1^* & -w & -w & -w \\
			-w & -2 x_2^* & 0 & 0 \\
			-w & 0 & -2 x_2^* & 0 \\
			-w &0 & 0 & -2 x_2^*
		\end{pmatrix}.
	\end{equation}
To exploit the symmetric network structure, we assume that $\sigma_2 = \sigma_3 = \sigma_4$.
Then, the Lyapunov equation reads
\begin{equation}
\begin{aligned}
    &\begin{pmatrix}
		-2 x_1^* & -w & -w & -w \\
		-w & -2 x_2^* & 0 & 0 \\
		-w & 0 & -2 x_2^* & 0 \\
		-w & 0 & 0 & -2 x_2^*
    \end{pmatrix}\begin{pmatrix}
        C_{11} & C_{12} & C_{13} & C_{14} \\
        C_{21} & C_{22} & C_{23} & C_{24} \\
        C_{31} & C_{32} & C_{33} & C_{34} \\
        C_{41} & C_{42} & C_{43} & C_{44}
    \end{pmatrix} \\  &+ \begin{pmatrix}
        C_{11} & C_{12} & C_{13} & C_{14} \\
        C_{21} & C_{22} & C_{23} & C_{24} \\
        C_{31} & C_{32} & C_{33} & C_{34} \\
        C_{41} & C_{42} & C_{43} & C_{44}
    \end{pmatrix} \begin{pmatrix}
		-2 x_1^* & -w & -w & -w \\
		-w & -2 x_2^* & 0 & 0 \\
		-w & 0 & -2 x_2^* & 0 \\
		-w & 0 & 0 & -2 x_2^*
    \end{pmatrix} = \begin{pmatrix}
        \sigma_1^2 & 0 & 0 & 0 \\
        0 & \sigma_2^2 & 0 & 0 \\
        0 & 0 & \sigma_2^2 & 0 \\
        0 & 0 & 0 & \sigma_2^2
    \end{pmatrix}.
\end{aligned}
\end{equation}
Because $C$ is symmetric, we consider the diagonal and upper triangular entries of $C$. By symmetry of the network, we obtain $C_{12} = C_{13} = C_{14}$, $C_{22} = C_{33} = C_{44}$, and $C_{23} = C_{24} = C_{34}$. Therefore, we only need to solve the following set of four linear equations for $C_{11}$, $C_{12}$, $C_{22}$, and $C_{23}$:
\begin{align}
    -2x_1^*C_{11} - wC_{12} - wC_{12} - wC_{12} - 2x_1^*C_{11} - wC_{12} - wC_{12} - wC_{12} &= \sigma_1^2, \label{C2-1st} \\
    -2x_1^*C_{12} - wC_{22} - wC_{23} - wC_{23} - wC_{11} - 2x_2^*C_{12} &= 0, \\
    -wC_{12} - 2x_2^*C_{22} - wC_{12} - 2x_2^*C_{22} &= \sigma_2^2, \\
    -wC_{13} - 2x_2^*C_{23} - wC_{12} - 2x_2^*C_{23} \label{C2-4th} &= 0.
\end{align}

\subsection{Lollipop}\label{c3lollipop}

The system of stochastic differential equations for the lollipop graph is given by
 \begin{align}
		dx_1 &=[f(x_1) + w(x_2+ 1)]dt + \sigma_1dW_1, \\
		dx_2 &= [f(x_2) + w(x_1+ 1) + w(x_3 + 1) + w(x_4 +1)]dt + \sigma_2dW_2, \\
		dx_3 &=[f(x_3) + w(x_2 + 1) + w(x_4 + 1)]dt + \sigma_3dW_3, \\
		dx_4 &= [f(x_4) + w(x_2+1) + w(x_3 + 1)]dt + \sigma_4dW_4.
 \end{align}
Owing to the symmetry of the network structure, we obtain $x_3^* = x_4^*$ and
 \begin{align}	
		(x_1^*)^2 +r + w(x_2^* + 1) &= 0, \label{first-triangle-line} \\
		(x_2^*)^2 + r + w(x_1^* + 1) + 2w(x_3^* + 1) &= 0, \label{second-lollipop} \\
		(x_3^*)^2 + r + w(x_2^* + 1)+ w(x_3^* + 1) &= 0. \label{third-lollipop}
 \end{align}
From Eq.~\eqref{first-triangle-line}, we deduce
\begin{equation}
x_2^* = \frac{-(x_1^*)^2 - r}{w} - 1.
\label{eq:x_2^*-lollipop}
\end{equation}
By substituting Eq.~\eqref{eq:x_2^*-lollipop} in Eqs.~\eqref{second-lollipop} and \eqref{third-lollipop}, we obtain
\begin{equation}\label{first equation}
	\left[ \frac{-(x_1^*)^2 - r}{w} - 1 \right]^2 + r + w(x_1^* + 1) + 2w(x_3^* + 1) = 0
\end{equation}
and 
\begin{equation}\label{second equation}
	(x_3^*)^2 - (x_1^*)^2 + w(x_3^* + 1) = 0.
\end{equation}
Note that Eq.~\eqref{second equation} represents two hyperbolas that do not depend on $r$ and that Eq.~\eqref{first equation} is a fourth degree polynomial in terms of $x_1^*$. As $r$ increases from $-1$, the number of intersections between the fourth-degree polynomial and the hyperbolas changes from eight to six, and then to four, two, and zero. We define $r_c'''$ to be the $r$ value at which the number of intersections changes from eight to six, $r_c''$ to be where it changes from six to four, $r_c'$ to be where it changes from four to two, and $r_c$ be where it changes from two to zero. Then, we obtain $r_c''' \approx -0.198$, $r_c''\approx -0.155$, $r_c' \approx -0.142$, and $r_c \approx -0.120$. Therefore, we select the two $r$ values $r=-0.2$ and $r=-0.4$ for calculating $d$.

The sign-flipped Jacobian in an equilibrium is 
\begin{equation}
		A = \begin{pmatrix}
			-2 x_1^* & -w & 0 & 0 \\
			-w & -2 x_2^* & -w & -w\\
			0 & -w & -2 x_3^* & -w \\
			0 & -w & -w & -2 x_3^*
		\end{pmatrix}.
	\end{equation}
To exploit the symmetric network structure, we assume that $\sigma_3 = \sigma_4$. Then, the Lyapunov equation reads
\begin{equation}
\begin{aligned}
    &\begin{pmatrix}
		-2 x_1^* & -w & 0 & 0 \\
		-w & -2 x_2^* & -w & -w \\
		0 & -w & -2 x_3^* & -w \\
		0 & -w & -w & -2 x_3^*
    \end{pmatrix}\begin{pmatrix}
        C_{11} & C_{12} & C_{13} & C_{14} \\
        C_{21} & C_{22} & C_{23} & C_{24} \\
        C_{31} & C_{32} & C_{33} & C_{34} \\
        C_{41} & C_{42} & C_{43} & C_{44}
    \end{pmatrix} \\  &+ \begin{pmatrix}
        C_{11} & C_{12} & C_{13} & C_{14} \\
        C_{21} & C_{22} & C_{23} & C_{24} \\
        C_{31} & C_{32} & C_{33} & C_{34} \\
        C_{41} & C_{42} & C_{43} & C_{44}
    \end{pmatrix} \begin{pmatrix}
		-2 x_1^* & -w & 0 & 0 \\
		-w & -2 x_2^* & -w & -w \\
		0 & -w & -2 x_3^* & -w \\
		0 & -w & -w & -2 x_3^*
    \end{pmatrix} = \begin{pmatrix}
        \sigma_1^2 & 0 & 0 & 0 \\
        0 & \sigma_2^2 & 0 & 0 \\
        0 & 0 & \sigma_3^2 & 0 \\
        0 & 0 & 0 & \sigma_3^2
    \end{pmatrix}.
\end{aligned}
\end{equation}
We consider the diagonal and upper triangular entries of $C$. By symmetry of the network, we obtain $C_{23} = C_{24}$, $C_{13} = C_{14}$, and $C_{33} = C_{44}$. Therefore, we need to solve the following set of seven linear equations for $C_{11}$, $C_{12}$, $C_{13}$, $C_{22}$, $C_{23}$, $C_{33}$, and $C_{34}$:
\begin{align}
    -2x_1^*C_{11} - wC_{12} - 2x_1^*C_{11} - wC_{12} &= \sigma_1^2, \label{c3-1st} \\
    -2x_1^*C_{12} - wC_{22}-wC_{11} - 2x_2^*C_{12} - wC_{13} - wC_{13} &= 0,\\
    -2x_1^*C_{13} - wC_{23} - wC_{12} - 2x_3^*C_{13} - wC_{13} &= 0, \\
    -wC_{12} - 2x_2^*C_{22} - wC_{23} - wC_{23} - wC_{12} -2x_2^*C_{22}- wC_{23} - wC_{23} &= \sigma_2^2, \\
    -wC_{13} - 2x_2^*C_{23} - wC_{33} - wC_{34} - wC_{22} -2x_3^*C_{23} - wC_{23} &= 0, \\
    -wC_{23} - 2x_3^*C_{33} - wC_{34} - wC_{23} - 2x_3^*C_{33} - wC_{34} &= \sigma_3^2, \\
    -wC_{23} - wC_{34} - 2x_3^*C_{33} - wC_{23} - wC_{33} - 2x_3^*C_{34} &= 0. \label{c3-7th}
\end{align}

\newpage

\section{Proof of the theorems}\label{sec: Appendix F}

\subsection{Proof of Theorem 1}

Without loss of generality, we calculate the CV of the EWS obtained by the ``diagonal" method with entries $\{(1, 1),$
$(2, 2)\}$ and the CV for the ``row" method with entries $\{(1, 1), (1, 2)\}$. The CV is defined to be the standard deviation divided by mean. Therefore, using Eqs.~(3.7) and (3.8) in the main text, we obtain the CV with entries $\{(1, 1), (2,2)\}$ as follows:
\begin{equation}\label{CV1122}
    \text{CV}_{\text{diagonal}} \equiv \sqrt{\frac{2}{L-1}}\frac{\sqrt{(C_{11})^2 + (C_{12})^2 + (C_{21})^2 + (C_{22})^2}}{C_{11} + C_{22}}.
\end{equation}
The CV with $\{(1, 1), (1, 2)\}$ is 
\begin{equation}\label{CV1112}
     \text{CV}_{\text{row}} \equiv  \sqrt{\frac{1}{L-1}}\frac{\sqrt{C_{11}C_{11} + C_{11}C_{12} + C_{11}C_{21} + C_{11}C_{12} + C_{11}C_{11} + C_{12}C_{11} + C_{11}C_{12} + C_{12}C_{12}}}{C_{11} + C_{12}}.
\end{equation}

By applying $C_{12} = C_{21}$ to Eq.~\eqref{CV1122}, we obtain
\begin{equation}\label{eq 85}
    (L-1) \left(\text{CV}_{\text{diagonal}}\right)^2 = \frac{2\left[(C_{11})^2 + 2(C_{12})^2 + (C_{22})^2\right]}{(C_{11} + C_{22})^2} = 2 + \frac{4(C_{12})^2 - 4 C_{11}C_{22} }{(C_{11} + C_{22})^2}.
\end{equation}
Similarly, we obtain
\begin{equation}\label{eq 86}
    (L-1) \left(\text{CV}_{\text{row}}\right)^2 = \frac{2(C_{11})^2 + 5C_{11}C_{12} + (C_{12})^2}{(C_{11} + C_{12})^2} = 2 + \frac{C_{11}C_{12} - (C_{12})^2}{(C_{11}+ C_{12})^2}.
\end{equation}
Because $(C_{12})^2 \leq C_{11}C_{22}$, Eqs.~\eqref{eq 85} and \eqref{eq 86} imply that
\begin{equation}
(L-1) \left(\text{CV}_{\text{diagonal}}\right)^2 \le 2 \le  (L-1) \left(\text{CV}_{\text{row}}\right)^2.
\end{equation}
Therefore, $\text{CV}_{\text{diagonal}} \le \text{CV}_{\text{row}}$.

\subsection{Proof of Theorem 2}\label{sec: Appendix G}

Without loss of generality, we calculate the CV of the EWS obtained by the ``diagonal" method with entries $\{(1, 1), (2, 2),$
$ \ldots, (n, n)\}$, denoted again by $\text{CV}_{\text{diagonal}}$, and the CV for the ``minor" method with entries $\{(i, j) : 1 \leq i, j \leq n\}$, denoted by $\text{CV}_{\text{minor}}$. We obtain
\begin{equation}\label{eq 87}
\text{CV}_{\text{diagonal}} = \sqrt{\frac{2}{L-1}}\frac{\sqrt{\sum_{i = 1}^n\sum_{j = 1}^n \left(C_{ij}\right)^2}}{\sum_{i = 1}^n C_{ii}},
\end{equation}
which yields 
\begin{equation}\label{eq 88}
(L-1) \left(\text{CV}_{\text{diagonal}}\right)^2 = 2\frac{\sum_{i = 1}^n \sum_{j = 1}^n (C_{ij})^2}{\left(\sum_{i = 1}^n C_{ii}\right)^2} \leq 2. 
\end{equation}
The inequality in Eq.~\eqref{eq 88} holds true because
\begin{equation}
    \left(\sum_{i = 1}^n C_{ii}\right)^2 - \sum_{i = 1}^n\sum_{j = 1}^n (C_{ij})^2 =  \sum_{i,j=1}^n C_{ii}C_{jj} - \sum_{i, j=1}^n (C_{ij})^2 \geq 0,
\end{equation}
which follows from $(C_{ij})^2 \leq C_{ii}C_{jj}$, $\forall i, \forall j \in \{1, \ldots, n\}$.

On the other hand, we obtain
\begin{equation}\label{eq 90}
(L-1) \left(\text{CV}_{\text{minor}}\right)^2 =
\frac{\sum_{m,m'=1}^{n^2} C_{i_m, i_{m'}}C_{j_m, j_{m'}} + \sum_{m, m'=1}^{n^2} C_{i_m, j_{m'}}C_{i_{m'}, j_m}}{\left(\sum_{m = 1}^{n^2}C_{i_m, j_m}\right)^2}
\end{equation}
where
$(i_1, j_1) = (1, 1)$, $\ldots$, $(i_n, j_n) = (1, n)$, $(i_{n+1}, j_{n+1}) = (2, 1)$, $\ldots$, $(i_{2n}, j_{2n}) = (2, n)$, $\ldots$, $(i_{n(n-1) + 1}, j_{n(n-1) + 1}) = (n, 1)$, $\ldots$, $(i_{n^2}, j_{n^2}) = (n, n)$.
Notice that then
\begin{equation}
    \sum_{m,m'=1}^{n^2} C_{i_m, i_{m'}}C_{j_m, j_{m'}} = \sum_{m, m'=1}^{n^2} C_{i_m, j_{m'}}C_{i_{m'}, j_m} = \left(\sum_{m = 1}^{n^2} C_{i_m, j_m}\right)^2.
\label{eq:minor-2}
\end{equation}
By substituting Eq.~\eqref{eq:minor-2} in Eq.~\eqref{eq 90}, we obtain
\begin{equation}
(L-1) \left(\text{CV}_{\text{minor}}\right)^2 = 2.
\label{eq:minor-final}
\end{equation}
By combining Eqs.~\eqref{eq 88} and \eqref{eq:minor-final}, we obtain $\text{CV}_{\text{diagonal}} \le \text{CV}_{\text{minor}}$.

\newpage

\section{Networks}\label{sec: Appendix E}

We use 23 networks used in our previous study \cite{maclaren2023early}.
We show the networks used in this study in Table \ref{Networks}. All empirical networks are from the ``networkdata" R package \cite{schoch2022}. We coerced all networks to become undirected, unweighted, and simple. If the obtained network is not connected, we used the largest connected component.

Here, we detail how we generated the five model networks. We generated the Erd\H{o}s-R\'{e}nyi (ER) network with $N = 100$ nodes and the probability of an edge between each pair of nodes equal to $0.05$.

We generated the ER islands by first grouping $N = 100$ nodes into five sub-networks of $20$ nodes each. The nodes within each sub-network were adjacent to each other with probability $0.2475$. Each pair of sub-networks are connected by exactly one edge, whose endpoints are selected uniformly randomly, one for each of the two sub-networks. Because the generated network had one isolated node, we removed it to consider the remainder of the entire network, which is the largest connected component.

We used a network generated by the Barab\'{a}si-Albert (BA) model with $N = 100$ and $m=2$ \cite{barabasi1999emergence}, where $m$ is the number of edges to which a new node connects when added to the network. The initial network for the BA model is a network with two nodes connected by an edge.

We configured the degree distribution of the fitness network to be a power-law distribution with power-law exponent $-2$. To this end, we set the fitness of node $i$ (with $i \in \{1, \ldots, N \}$) to $(i + i_0 - 1)^{-\alpha}$, where $i_0 = N^{1 - \frac{1}{\alpha}}\left[ 10\sqrt{2}(1-\alpha) \right]^{\frac{1}{\alpha}}$ and $\alpha = 1$ \cite{cho2009percolation}. We used $N = 100$ nodes to generate the fitness network. Because 13 nodes were isolated, the largest connected component had $87$ nodes.

We also generated a network with power-law degree distribution using a configuration model. We let $N = 100$ and sample the degree of each node using the Pareto distribution: a node has degree $k \in \{1, 2, \ldots\}$ with probability $p(k) = \frac{k^{-\alpha}}{\sum_{k'=k_{\min}}^{N-1} (k')^{-\alpha}}$ with $k \in \{k_{\min}, \ldots, N-1\}$, $\alpha = -2$, and $k_{\min} = 1$. We used an algorithm by Viger and Latapy \cite{csardi2006igraph}, which samples all possible simple undirected graphs conditioned on a given input degree distribution and the network being connected.

Finally, we generated a Lancichinetti-Fortunato-Radicchi (LFR) network with $100$ nodes. We set the power-law exponent of the expected degree distribution and the community size distribution to $-2$ and $-1.5$, respectively. We also set the average degree to $5$, maximum degree to $20$, maximum community size to $20$, minimum community size to $10$, and the probability that an edge connects nodes in different communities to $0.1$. We used NetworkX (v2.7.1) package \cite{hagberg2008exploring} for Python (v3.10.2) to generate this network.

\begin{longtable}{|ccccc|}
	\caption{Networks used in the present study. $\left| E \right|$: number of edges. $\langle k\rangle$: average degree. ER: Erd\H{o}s-R\'{e}nyi. BA: Barab\'{a}si-Albert. LFR: Lancichinetti-Fortunato-Radicchi.}
	\label{Networks}
	\endfirsthead
	\endhead
	\endfoot
	\hline
	Network & $N$ & {$\left| E \right|$} & {$\langle k\rangle$} & Reference\\
	\hline
	Karate & 34 & 78 & 4.59 & \cite{zachary1977information} \\
	Bat & 43 & 546 & 25.40 & \cite{silvis2014roosting} \\
	Surfer & 43 & 336 & 15.63 & \cite{freeman1988human} \\
	Elephant seal & 46 & 56 & 2.44 & \cite{casey2015rival} \\
	Lizard & 60 & 318 & 10.60 & \cite{bull2012social} \\
	Dolphin & 62 & 159 & 5.13 & \cite{lusseau2003bottlenose} \\
	Weaver bird & 64 & 177 & 5.53 & \cite{van2014cooperative} \\
	Highschool boy & 70 & 274 & 7.83 & \cite{coleman1964introduction} \\
	Tortoise & 94 & 181 & 3.85 & \cite{sah2016inferring} \\
	House finch & 108 & 1026 & 16.03 & \cite{adelman2015feeder} \\
	Vole & 111 & 240 & 4.32 & \cite{davis2015spatial} \\
	Nestbox & 126 & 1615 & 25.64 & \cite{firth2015experimental} \\
	Pira & 128 & 201 & 3.14 & \cite{gill2014lethal} \\
	Drug user & 193 & 273 & 2.83 & \cite{weeks2002social}\\
	Jazz & 198 & 2742 & 27.70 & \cite{gleiser2003community} \\
	Hall & 217 & 1839 & 16.95 & \cite{freeman1998exploring}\\
	Netsci & 379 & 914 & 4.82 & \cite{newman2006finding} \\
        ER & 100 & 246 & 4.92 & \cite{erdo1959s} \\
	ER islands & 99 & 244 & 4.93 & \cite{csardi2006igraph} \\
	BA & 100 & 197 & 3.94 & \cite{barabasi1999emergence} \\
    	Fitness & 87 & 197 & 4.53 & \cite{goh2001universal} \\
	Configuration & 100 & 146 & 2.92 & \cite{clauset2009power}\\
	LFR & 100 & 269 & 5.38 & \cite{lancichinetti2008benchmark} \\
	\hline
\end{longtable}

\newpage

\section{Comparison between the dominant eigenvalue method and our EWSs}\label{dominant eigenvalue}

The dominant eigenvalue of the sample covariance matrix depends on all the entries of the matrix and is an often employed EWS.
Because our EWSs only use entries of the principal minor matrix composed of the selected sentinel nodes, 
we compare the dominant eigenvalue of the principal minor of the sample covariance matrix with our EWSs in this section.

\begin{figure}[ht]
    \centering
    \includegraphics[width=0.97\linewidth]{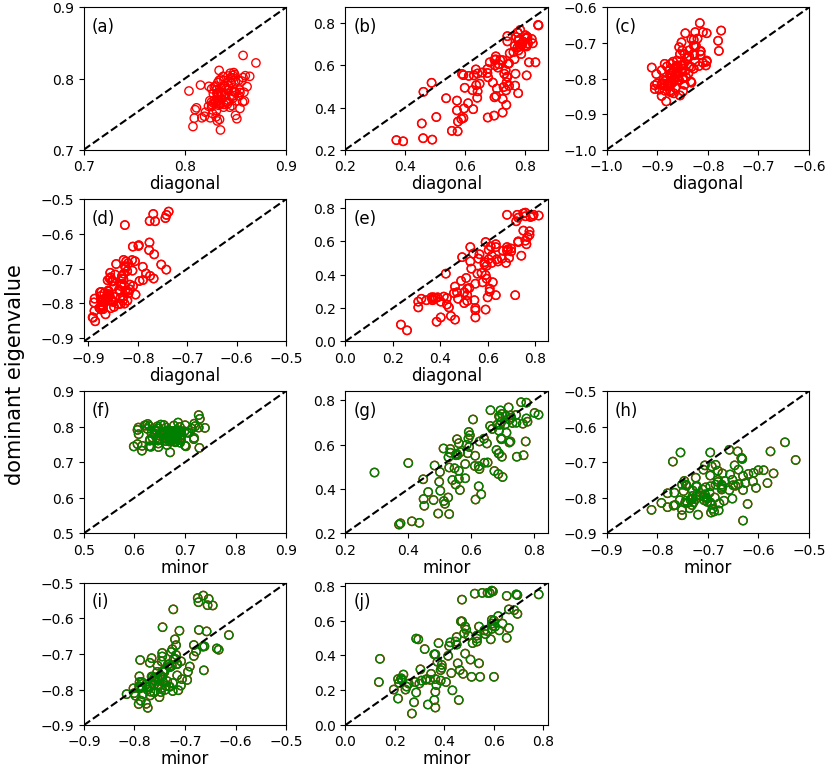}
    \caption{
      Comparison between the Kendall's $\tau$ value for the dominant eigenvalue method and that for our proposed EWSs. (a)--(e): Comparison between the dominant eigenvalue method and the ``diagonal" method.
(f)--(j) Comparison between the dominant eigenvalue method and the ``minor'' method.    
(a) and (f): Coupled double-well dynamics on the BA network with 100 nodes, with homogeneous stress and noise, $u$ as the control parameter, and ascending simulations. (b) and (g): SIS dynamics on the dolphin network, with heterogeneous noise, $D$ as the control parameter, and ascending simulations. (c) and (h): Coupled double-well dynamics on the lizard network, with homogeneous stress and noise, $D$ as the control parameter, and descending simulations. (d) and (i): Mutualistic interaction dynamics on the Netsci network, with heterogeneous stress and noise, $D$ as the control parameter, and descending simulations. (e) and (j): Coupled double-well dynamics on the surfers network, with heterogeneous stress and noise, with $D$ as the control parameter, and ascending simulations.
Each circle represents the $\tau$ values for a sentinel node set with five nodes selected uniformly at random.
The dashed lines represent the diagonal.}
    \label{fig:dominant-eigenvalue}
\end{figure}

In Fig.~\ref{fig:dominant-eigenvalue}(a)--(e), we compare the Kendall's $\tau$ between the dominant eigenvalue method and our ``diagonal'' method for the five pairs of the dynamics and networks used in Figs. 3 and 4 of the main text in the order they appear there (i.e., Fig. 3(a), 3(b), 4(a), 4(b), and 4(c)). We find that the dominant eigenvalue method is inferior to the ``diagonal'' method.
It should be noted that a larger absolute value of $\tau$ is better in all the panels in Fig.~\ref{fig:dominant-eigenvalue} because the simulations are in the ascending direction in 
Fig.~\ref{fig:dominant-eigenvalue}(a), (b), and (e) (i.e., below the diagonal is better), and in the descending direction in Fig.~\ref{fig:dominant-eigenvalue}(c) and (d) (i.e., above the diagonal is better). This result further strengthens our main conclusion that the use of off-diagonal entries of the covariance matrix deteriorates the quality of EWSs.

The dominant eigenvalue method and our ``minor'' method both use all the entries of the principal minor of the sample covariance matrix.
Therefore, we compare $\tau$ between the dominant eigenvalue method and the ``minor'' method in Fig.~\ref{fig:dominant-eigenvalue}(f)--(j) for the same five pairs of the dynamics and networks. We observe that which of the two EWSs is better than the other depends on the pair of dynamics and network. This result does not compromise our main conclusion.

\newpage
\clearpage

\section{Multivariate linear regressions for explaining $\tau$ and $d$}\label{sec: Appendix D}

To statistically investigate the performance of the different types of EWS across the dynamics models, networks, and other factors, we run multivariate linear regressions on our numerical results.
For each dynamical system model, we set the dependent variable to either $\tau$ or $d$. 
There are up to five independent variables in each multiple linear regression: the network (one of the 23 networks; categorical), the method of selecting entry combinations of the covariance matrix (``diagonal'', ``minor'', ``row'', or ``random''; categorical), the control parameter ($D$ or $u$; categorical and binary; the SIS dynamics does not have this categorical independent variable because we only use $D$ as control parameter for the SIS model), whether the stress and noise given to the nodes are both homogeneous or both heterogeneous (categorical and binary), and direction of simulation (i.e., either ascending or descending; categorical and binary; only for the coupled double-well and SIS dynamics). Because we produce the entries of the sample covariance matrix for constructing an EWS, $S$, one hundred times for each combination of the independent variables, there are $100 \times 23 \times 4 \times 2 \times 2 = 36800$ sample points for the six regression models for the mutualistic interaction, gene regulatory, and SI dynamics, and $100 \times 23 \times 4 \times 2 \times 2 \times 2 = 73600$ sample points for the two regression models for the coupled double-well dynamics.

We test eight multivariate linear regressions, one per pair of the dynamics (i.e., coupled double-well, mutualistic interaction, gene regulatory, or SIS) and the dependent variable ($\tau$ or $d$).
We used the Python package statsmodels (v.0.14.4) to perform these multivariate linear regressions.
We show the results in Tables~\ref{Coupled double-well tau}--\ref{SIS d}.

\begin{longtable}{|l r >{\centering\arraybackslash}p{0.06\textwidth} >{\centering\arraybackslash}p{0.18\textwidth} > {\centering\arraybackslash}p{0.12\textwidth} > {\centering\arraybackslash}p{0.03\textwidth} > {\centering\arraybackslash}p{0.10\textwidth}|}
     \caption{Regression results for the coupled double-well dynamics when the dependent variable is $\tau$. SE: standard error. CI: confidence interval. Including in Tables~\ref{coupled double-well d}--\ref{SIS d}, we used the LFR network,
``diagonal'' method, $u$, heterogeneous, and descending as the reference for the network, EWS type, control parameter, homogeneity, and direction of simulation independent variables, respectively. We have obtained $R^2 = 0.779$ and $F = 9,274$.} 
    \label{Coupled double-well tau}
    \endfirsthead
     \endhead
     \endfoot
    \hline
     Variable & {Coefficient} & {SE} & CI & $p$ & $df$ & $F$ \\

    \hline
    Intercept & 0.599 & 0.003 & $[0.593, 0.605]$ & $< 10^{-7}$ & &  \\
    Network &  &  &  & $< 10^{-7}$ & 22 & 254.482 \\
        \hspace{2 em} Karate & 0.042 & 0.004 & $[0.034, 0.050]$ & $< 10^{-7}$ && \\
        \hspace{2 em} Bat & 0.104 & 0.004 & $[0.095, 0.111]$ & $< 10^{-7}$ & &  \\
        \hspace{2 em} Surfer & $0.082$ & 0.004 & $[0.074, 0.090]$ & $< 10^{-7}$&& \\
        \hspace{2 em} Elephant seal & 0.018 & 0.004 & $[0.010, 0.026]$ & $8.281 \times 10^{-6}$ & & \\
        \hspace{2 em} Lizard & 0.061 & 0.004 & $[0.053, 0.069]$ & $< 10^{-7}$ && \\
        \hspace{2 em} Dolphin & 0.054 & 0.004 & $[0.046, 0.062]$ & $< 10^{-7}$ & &  \\
        \hspace{2 em} Weaver bird & 0.016 & 0.004 & $[0.008, 0.024]$ & $1.018 \times 10^{-4}$&& \\
        \hspace{2 em} Highschool boy & 0.045 & 0.004 & $[0.037, 0.053]$ & $< 10^{-7}$ && \\
        \hspace{2 em} Tortoise & 0.009 & 0.004 & $[0.001, 0.017]$ & $0.033$&& \\
        \hspace{2 em} House finch & $0.014$ & 0.004 & $[0.006, 0.022]$ & $6.352 \times 10^{-4}$ && \\
        \hspace{2 em} Vole & 0.029 & 0.004 & $[0.021, 0.037]$ & $< 10^{-7}$&& \\
        \hspace{2 em} Nestbox & 0.014 & 0.004 & $[0.006, 0.022]$ & $4.191 \times 10^{-4}$ && \\
        \hspace{2 em} Pira & $-0.041$ & 0.004 & $[-0.049, -0.033]$ & $< 10^{-7}$&& \\
        \hspace{2 em} Drug user & $-0.029$ & 0.004 & $[-0.037, -0.021]$ & $< 10^{-7}$ & & \\
        \hspace{2 em} Jazz & $-0.043$ & 0.004 & $[-0.051, -0.035]$ & $< 10^{-7}$ && \\
        \hspace{2 em} Hall & $-0.006$ & 0.004 & $[-0.014, 0.002]$ & $0.158$ && \\
        \hspace{2 em} Netsci & $-0.046$ & 0.004 & $[-0.054, -0.038]$ & $< 10^{-7}$&& \\
        \hspace{2 em} ER & 0.040 & 0.004 & $[0.032, 0.048]$ & $< 10^{-7}$ && \\
        \hspace{2 em} ER islands & 0.052 & 0.004 & $[0.044, 0.060]$ & $< 10^{-7}$ & & \\
        \hspace{2 em} BA & $-0.020$ & 0.004 & $[-0.028, -0.012]$ & $8.094 \times 10^{-7}$ & &  \\
        \hspace{2 em} Fitness & $-0.052$ & 0.004 & $[-0.060, -0.044]$ & $< 10^{-7}$ && \\
        \hspace{2 em} Configuration & $-0.078$ & 0.004 & $[-0.086, -0.070]$ & $< 10^{-7}$&& \\
    Control parameter & & & & $<10^{-7}$ & 1 & 9,913.796 \\
    \hspace{2 em} $u$ & 0.120 & 0.001 & $[0.117, 0.122]$ & $< 10^{-7}$ && \\
    Stress and noise & & & & $<10^{-7}$ & 1 & 327.120 \\
    \hspace{2 em} heterogeneous & $-0.022$ & 0.001 & $[-0.024, -0.019]$ & $< 10^{-7}$ && \\
    Simulation direction & & & & $< 10^{-7}$ & 1 & 3,188.435 \\
    \hspace{2 em} descending & 0.068 & 0.001 & $[0.066, 0.070]$ & $< 10^{-7}$ && \\
    Method  & & & & $<10^{-7}$ & 3 & 80,214.738 \\
    \hspace{2 em} minor & $-0.154$ & 0.002 & $[-0.158, -0.151]$ & $< 10^{-7}$ && \\
    \hspace{2 em} row & $-0.666$ & 0.002 & $[-0.669, -0.663]$ & $< 10^{-7}$ && \\
    \hspace{2 em} random & $-0.648$ & 0.002 & $[-0.652, -0.645]$ & $< 10^{-7}$ && \\
    \hline
\end{longtable}

\newpage

\begin{longtable}{|l r >{\centering\arraybackslash}p{0.06\textwidth} >{\centering\arraybackslash}p{0.18\textwidth} > {\centering\arraybackslash}p{0.12\textwidth} > {\centering\arraybackslash}p{0.03\textwidth} > {\centering\arraybackslash}p{0.10\textwidth}|}
          \caption{Regression results for the coupled double-well dynamics when the dependent variable is $d$. We have obtained $R^2 = 0.618$ and $F = 4,242$.}
  \label{coupled double-well d}
  \endfirsthead
    \hline
     Variable & {Coefficient} & {SE} & CI & $p$ & $df$ & $F$ \\
    \hline
    Intercept & 3.968 & 0.027 & $[3.914, 4.021]$ & $< 10^{-7}$ & &  \\
    Network &  &  &  & $< 10^{-7}$ & 22 & 150.167 \\
        \hspace{2 em} Karate & 0.296 & 0.034 & $[0.230, 0.363]$ & $< 10^{-7}$ && \\
        \hspace{2 em} Bat & 0.748 & 0.034 & $[0.681, 0.814]$ & $< 10^{-7}$ & &  \\
        \hspace{2 em} Surfer & $0.526$ & 0.034 & $[0.459, 0.592]$ & $< 10^{-7}$&& \\
        \hspace{2 em} Elephant seal & 0.234 & 0.034 & $[0.167, 0.301]$ & $< 10^{-7}$ & & \\
        \hspace{2 em} Lizard & 0.371 & 0.034 & $[0.304, 0.438]$ & $< 10^{-7}$ && \\
        \hspace{2 em} Dolphin & 0.443 & 0.034 & $[0.376, 0.510]$ & $< 10^{-7}$ & &  \\
        \hspace{2 em} Weaver bird & 0.095 & 0.034 & $[0.028, 0.162]$ & $5.343 \times 10^{-3}$ && \\
        \hspace{2 em} Highschool boy & 0.324 & 0.034 & $[0.257, 0.390]$ & $< 10^{-7}$ && \\
        \hspace{2 em} Tortoise & 0.141 & 0.034 & $[0.074, 0.208]$ & $3.505 \times 10^{-5}$ && \\
        \hspace{2 em} House finch & $-0.034$ & 0.034 & $[-0.101, 0.033]$ & $0.317$ && \\
        \hspace{2 em} Vole & 0.264 & 0.034 & $[0.197, 0.330]$ & $< 10^{-7}$&& \\
        \hspace{2 em} Nestbox & 0.076 & 0.034 & $[0.010, 0.143]$ & 0.025 && \\
        \hspace{2 em} Pira & $-0.112$ & 0.034 & $[-0.179, -0.046]$ & $9.624 \times 10^{-4}$&& \\
        \hspace{2 em} Drug user & $-0.043$ & 0.034 & $[-0.110, 0.024]$ & $0.207$ & & \\
        \hspace{2 em} Jazz & $-0.364$ & 0.034 & $[-0.431, -0.297]$ & $< 10^{-7}$ && \\
        \hspace{2 em} Hall & $-0.176$ & 0.034 & $[-0.242, -0.109]$ & $2.586 \times 10^{-7}$ && \\
        \hspace{2 em} Netsci & $-0.229$ & 0.034 & $[-0.296, -0.162]$ & $< 10^{-7}$&& \\
        \hspace{2 em} ER & 0.319 & 0.034 & $[0.252, 0.386]$ & $< 10^{-7}$ && \\
        \hspace{2 em} ER islands & 0.452 & 0.034 & $[0.386, 0.519]$ & $< 10^{-7}$ & & \\
        \hspace{2 em} BA & $0.001$ & 0.034 & $[-0.066, 0.068]$ & 0.982 & &  \\
        \hspace{2 em} Fitness & $-0.269$ & 0.034 & $[-0.336, -0.202]$ & $< 10^{-7}$ && \\
        \hspace{2 em} Configuration & $-0.333$ & 0.034 & $[-0.400, -0.266]$ & $< 10^{-7}$&& \\
    Control parameter & & & & $<10^{-7}$ & 1 & 11,692.119 \\
    \hspace{2 em} $u$ & 1.086 & 0.010 & $[1.067, 1.106]$ & $< 10^{-7}$ && \\
    Stress and noise & & & & $<10^{-7}$ & 1 & 1,129.676 \\
    \hspace{2 em} heterogeneous & $-0.338$ & 0.010 & $[-0.357, -0.318]$ & $< 10^{-7}$ && \\
    Simulation direction & & & & $< 10^{-7}$ & 1 & 2,988.537 \\
    \hspace{2 em} descending & 0.549 & 0.010 & $[0.530, 0.569]$ & $< 10^{-7}$ && \\
    Method  & & & & $<10^{-7}$ & 3 & 33,222.597 \\
    \hspace{2 em} minor & $-2.246$ & 0.014 & $[-2.273, -2.218]$ & $< 10^{-7}$ && \\
    \hspace{2 em} row & $-3.898$ & 0.014 & $[-3.926, -3.871]$ & $< 10^{-7}$ && \\
    \hspace{2 em} random & $-3.838$ & 0.014 & $[-3.866, -3.810]$ & $< 10^{-7}$ && \\
    \hline
\end{longtable}

\newpage

\begin{longtable}{|l r >{\centering\arraybackslash}p{0.06\textwidth} >{\centering\arraybackslash}p{0.18\textwidth} > {\centering\arraybackslash}p{0.12\textwidth} > {\centering\arraybackslash}p{0.03\textwidth} > {\centering\arraybackslash}p{0.10\textwidth}|}
	\caption{Regression results for the mutualistic interaction dynamics when the dependent variable is $\tau$. We have obtained $R^2 = 0.782$ and $F = 4,883$.}
	\label{Mutualistic tau}
	\endfirsthead
	\endhead
	\endfoot
	\hline
	Variable & {Coefficient} & {SE} & CI & $p$ & $df$ & $F$ \\
	
	\hline
	Intercept & 0.808 & 0.004 & $[0.799, 0.817]$ & $< 10^{-7}$ & &  \\
	Network &  &  &  & $< 10^{-7}$ & 22 & 93.592 \\
	\hspace{2 em} Karate & 0.061 & 0.006 & $[0.050, 0.074]$ & $< 10^{-7}$ && \\
	\hspace{2 em} Bat & $-0.026$ & 0.006 & $[-0.037, -0.014]$ & $2.432 \times 10^{-5}$ & &  \\
	\hspace{2 em} Surfer & $0.062$ & 0.006 & $[0.050, 0.074]$ & $< 10^{-7}$&& \\
	\hspace{2 em} Elephant seal & 0.066 & 0.006 & $[0.054, 0.078]$ & $< 10^{-7}$ & & \\
	\hspace{2 em} Lizard & 0.008 & 0.006 & $[-0.004, 0.020]$ & $0.193$ && \\
	\hspace{2 em} Dolphin & 0.034 & 0.006 & $[0.023, 0.046]$ & $< 10^{-7}$ & &  \\
	\hspace{2 em} Weaver bird & 0.020 & 0.006 & $[0.008, 0.032]$ & $9.401 \times 10^{-4}$ && \\
	\hspace{2 em} Highschool boy & 0.023 & 0.006 & $[0.011, 0.035]$ & $1.772 \times 10^{-4}$ && \\
	\hspace{2 em} Tortoise & 0.030 & 0.006 & $[0.018, 0.042]$ & $9.081 \times 10^{-7}$&& \\
	\hspace{2 em} House finch & $-0.060$ & 0.006 & $[-0.072, -0.049]$ & $< 10^{-7}$ && \\
	\hspace{2 em} Vole & 0.016 & 0.006 & $[0.005, 0.028]$ & $0.006$&& \\
	\hspace{2 em} Nestbox & $-0.048$ & 0.006 & $[-0.059, -0.036]$ & $< 10^{-7}$ && \\
	\hspace{2 em} Pira & $0.023$ & 0.006 & $[0.011, 0.035]$ & $1.281 \times 10^{-4}$&& \\
	\hspace{2 em} Drug user & $0.024$ & 0.006 & $[0.012, 0.036]$ & $5.829 \times 10^{-5}$ & & \\
	\hspace{2 em} Jazz & $-0.089$ & 0.006 & $[-0.101, -0.077]$ & $< 10^{-7}$ && \\
	\hspace{2 em} Hall & $-0.064$ & 0.006 & $[-0.076, -0.053]$ & $< 10^{-7}$ && \\
	\hspace{2 em} Netsci & $-0.009$ & 0.006 & $[-0.021, 0.002]$ & $0.121$&& \\
    	\hspace{2 em} ER & $-0.009$ & 0.006 & $[-0.021, 0.003]$ & $0.152$ && \\
	\hspace{2 em} ER islands & 0.012 & 0.006 & $[-0.000, 0.024]$ & $0.054$ & & \\
	\hspace{2 em} BA & $0.053$ & 0.006 & $[0.041, 0.065]$ & $< 10^{-7}$ & &  \\
    \hspace{2 em} Fitness & $0.013$ & 0.006 & $[0.001, 0.025]$ & $0.037$ && \\
	\hspace{2 em} Configuration & $0.026$ & 0.006 & $[0.015, 0.038]$ & $1.291 \times 10^{-5}$&& \\
	Control parameter & & & & $<10^{-7}$ & 1 & 8,733.819 \\
	\hspace{2 em} $u$ & $-0.167$ & 0.002 & $[-0.171, -0.164]$ & $< 10^{-7}$ && \\
	Stress and noise & & & & $<10^{-7}$ & 1 & 138.574 \\
	\hspace{2 em} heterogeneous & $-0.021$ & 0.002 & $[-0.025, -0.018]$ & $< 10^{-7}$ && \\ 
	Method  & & & & $<10^{-7}$ & 3 & 40,301.292 \\
	\hspace{2 em} minor & $-0.133$ & 0.003 & $[-0.138, -0.128]$ & $< 10^{-7}$ && \\
	\hspace{2 em} row & $-0.692$ & 0.003 & $[-0.697, -0.687]$ & $< 10^{-7}$ && \\
	\hspace{2 em} random & $-0.669$ & 0.003 & $[-0.674, -0.664]$ & $< 10^{-7}$ && \\
	\hline
\end{longtable}

\newpage

\begin{longtable}{|l r >{\centering\arraybackslash}p{0.06\textwidth} >{\centering\arraybackslash}p{0.18\textwidth} > {\centering\arraybackslash}p{0.12\textwidth} > {\centering\arraybackslash}p{0.03\textwidth} > {\centering\arraybackslash}p{0.10\textwidth}|}
	\caption{Regression results for the mutualistic interaction dynamics when the dependent variable is $d$. We have obtained $R^2 = 0.643$ and $F = 2,451$.}
	\label{Mutualistic d}
	\endfirsthead
	\endhead
	\endfoot
	\hline
	Variable & {Coefficient} & {SE} & CI & $p$ & $df$ & $F$ \\
	
	\hline
	Intercept & 6.716 & 0.045 & $[6.627, 6.805]$ & $< 10^{-7}$ & &  \\
	Network &  &  &  & $< 10^{-7}$ & 22 & 22.311 \\
	\hspace{2 em} Karate & 0.045 & 0.058 & $[-0.069, 0.159]$ & $0.436$ && \\
	\hspace{2 em} Bat & 0.388 & 0.058 & $[0.274, 0.502]$ & $< 10^{-7}$ & &  \\
	\hspace{2 em} Surfer & $0.659$ & 0.058 & $[0.545, 0.773]$ & $< 10^{-7}$&& \\
	\hspace{2 em} Elephant seal & 0.177 & 0.058 & $[0.063, 0.292]$ & $3.566 \times 10^{-4}$ & & \\
	\hspace{2 em} Lizard & 0.208 & 0.058 & $[0.094, 0.322]$ & $0.010$ && \\
	\hspace{2 em} Dolphin & 0.249 & 0.058 & $[0.135, 0.363]$ & $1.881 \times 10^{-5}$ & &  \\
	\hspace{2 em} Weaver bird & 0.031 & 0.058 & $[-0.083, 0.145]$ & 0.598 && \\
	\hspace{2 em} Highschool boy & 0.312 & 0.058 & $[0.198, 0.426]$ & $< 10^{-7}$ && \\
	\hspace{2 em} Tortoise & 0.078 & 0.058 & $[-0.036, 0.192]$ & $0.181$&& \\
	\hspace{2 em} House finch & $0.107$ & 0.058 & $[-0.007, 0.221]$ & 0.067 && \\
	\hspace{2 em} Vole & 0.146 & 0.058 & $[0.032, 0.260]$ & $0.012$&& \\
	\hspace{2 em} Nestbox & 0.150 & 0.058 & $[0.036, 0.265]$ & $0.010$ && \\
	\hspace{2 em} Pira & $-0.148$ & 0.058 & $[-0.263, -0.034]$ & $0.011$&& \\
	\hspace{2 em} Drug user & $-0.049$ & 0.058 & $[-0.163, 0.065]$ & $0.398$ & & \\
	\hspace{2 em} Jazz & $0.029$ & 0.058 & $[-0.085, 0.144]$ & $0.436$ && \\
	\hspace{2 em} Hall & $0.140$ & 0.058 & $[0.026, 0.254]$ & $0.016$ && \\
	\hspace{2 em} Netsci & $-0.182$ & 0.058 & $[-0.297, -0.068]$ & $0.002$&& \\
    \hspace{2 em} ER & $-0.005$ & 0.058 & $[-0.119, 0.109]$ & $0.932$ && \\
	\hspace{2 em} ER islands & 0.154 & 0.058 & $[0.040, 0.269]$ & $0.008$ & & \\
	\hspace{2 em} BA & $0.247$ & 0.058 & $[0.133, 0.361]$ & $2.236 \times 10^{-5}$ & &  \\
    \hspace{2 em} Fitness & $-0.132$ & 0.058 & $[-0.246, -0.017]$ & $0.024$ && \\
	\hspace{2 em} Configuration & $-0.090$ & 0.058 & $[-0.204, 0.024]$ & $0.053$&& \\
	Control parameter & & & & $<10^{-7}$ & 1 & 13,306.851 \\
	\hspace{2 em} $u$ & $-1.981$ & 0.017 & $[-2.014, -1.947]$ & $< 10^{-7}$ && \\
	Stress and noise & & & & $<10^{-7}$ & 1 & 542.720 \\
	\hspace{2 em} heterogeneous & $-0.400$ & 0.017 & $[-0.434, -0.366]$ & $< 10^{-7}$ && \\ 
	Method  & & & & $<10^{-7}$ & 3 & 17,280.903 \\
	\hspace{2 em} minor & $-2.617$ & 0.024 & $[-2.665, -2.569]$ & $< 10^{-7}$ && \\
	\hspace{2 em} row & $-4.788$ & 0.024 & $[-4.836, -4.741]$ & $< 10^{-7}$ && \\
	\hspace{2 em} random & $-4.716$ & 0.024 & $[-4.764, -4.669]$ & $< 10^{-7}$ && \\
	\hline
\end{longtable}

\newpage

\begin{longtable}{|l r >{\centering\arraybackslash}p{0.06\textwidth} >{\centering\arraybackslash}p{0.18\textwidth} > {\centering\arraybackslash}p{0.12\textwidth} > {\centering\arraybackslash}p{0.03\textwidth} > {\centering\arraybackslash}p{0.10\textwidth}|}
	\caption{Regression results for the gene regulatory dynamics when the dependent variable is $\tau$. We have obtained $R^2 = 0.151$ and $F = 241.6$.}
	\label{Gene tau}
	\endfirsthead
	\endhead
	\endfoot
	\hline
	Variable & {Coefficient} & {SE} & CI & $p$ & $df$ & $F$ \\
	
	\hline
	Intercept & 0.117 & 0.004 & $[0.110, 0.124]$ & $< 10^{-7}$ & &  \\
	Network &  &  &  & $< 10^{-7}$ & 22 & 157.081 \\
	\hspace{2 em} Karate & 0.037 & 0.005 & $[0.028, 0.046]$ & $< 10^{-7}$ && \\
	\hspace{2 em} Bat & $-0.047$ & 0.005 & $[-0.056, -0.038]$ & $< 10^{-7}$ & &  \\
	\hspace{2 em} Surfer & $0.015$ & 0.005 & $[0.005, 0.024]$ & $0.002$ && \\
	\hspace{2 em} Elephant seal & 0.077 & 0.005 & $[0.068, 0.086]$ & $< 10^{-7}$ & & \\
	\hspace{2 em} Lizard & $-0.028$ & 0.005 & $[-0.037, -0.019]$ & $< 10^{-7}$ && \\
	\hspace{2 em} Dolphin & $-0.003$ & 0.005 & $[-0.012, 0.006]$ & $0.548$ & &  \\
	\hspace{2 em} Weaver bird & $-0.012$ & 0.005 & $[-0.021, -0.002]$ & 0.013 && \\
	\hspace{2 em} Highschool boy & 0.000 & 0.005 & $[-0.009, 0.009]$ & $0.959$ && \\
	\hspace{2 em} Tortoise & 0.031 & 0.005 & $[0.022, 0.040]$ & $< 10^{-7}$&& \\
	\hspace{2 em} House finch & $-0.059$ & 0.005 & $[-0.068, -0.050]$ & $< 10^{-7}$ && \\
	\hspace{2 em} Vole & $0.003$ & 0.005 & $[-0.006, 0.012]$ & $0.484$&& \\
	\hspace{2 em} Nestbox & $-0.045$ & 0.005 & $[-0.054, -0.036]$ & $< 10^{-7}$ && \\
	\hspace{2 em} Pira & $0.012$ & 0.005 & $[0.003, 0.021]$ & $0.011$ && \\
	\hspace{2 em} Drug user & $0.007$ & 0.005 & $[-0.003, 0.016]$ & $0.155$ & & \\
	\hspace{2 em} Jazz & $-0.058$ & 0.005 & $[-0.067, -0.049]$ & $< 10^{-7}$ && \\
	\hspace{2 em} Hall & $-0.053$ & 0.005 & $[-0.062, -0.044]$ & $< 10^{-7}$ && \\
	\hspace{2 em} Netsci & $-0.008$ & 0.005 & $[-0.017, 0.002]$ & $0.103$&& \\
    \hspace{2 em} ER & $-0.059$ & 0.005 & $[-0.068, -0.050]$ & $< 10^{-7}$ && \\
	\hspace{2 em} ER islands & $-0.014$ & 0.005 & $[-0.023, -0.005]$ & $0.003$ & & \\
	\hspace{2 em} BA & $0.094$ & 0.005 & $[0.085, 0.103]$ & $< 10^{-7}$ & &  \\
    	\hspace{2 em} Fitness & $-0.026$ & 0.005 & $[-0.035, -0.017]$ & $< 10^{-7}$ && \\
	\hspace{2 em} Configuration & $-0.048$ & 0.005 & $[-0.057, -0.038]$ & $< 10^{-7}$&& \\
	Control parameter & & & & $<10^{-7}$ & 1 & 47.089 \\
	\hspace{2 em} $u$ & $0.009$ & 0.001 & $[0.007, 0.012]$ & $< 10^{-7}$ && \\
	Stress and noise & & & & $2.945 \times 10^{-5}$ & 1 & 17.457 \\
	\hspace{2 em} heterogeneous & $-0.006$ & 0.001 & $[-0.08, -0.003]$ & $2.945 \times 10^{-5}$ && \\ 
	Method  & & & & $<10^{-7}$ & 3 & 1,001.198 \\
	\hspace{2 em} minor & $-0.034$ & 0.002 & $[-0.037, -0.030]$ & $< 10^{-7}$ && \\
	\hspace{2 em} row & $-0.089$ & 0.002 & $[-0.093, -0.085]$ & $< 10^{-7}$ && \\
	\hspace{2 em} random & $-0.087$ & 0.002 & $[-0.091, -0.083]$ & $< 10^{-7}$ && \\
	\hline
\end{longtable}

\newpage

\begin{longtable}{|l r >{\centering\arraybackslash}p{0.06\textwidth} >{\centering\arraybackslash}p{0.18\textwidth} > {\centering\arraybackslash}p{0.12\textwidth} > {\centering\arraybackslash}p{0.03\textwidth} > {\centering\arraybackslash}p{0.10\textwidth}|}
	\caption{Regression results for the gene regulatory dynamics when the dependent variable is $d$. We have obtained $R^2 = 0.034$ and $F = 48.33$.}
	\label{Gene d}
	\endfirsthead
	\endhead
	\endfoot
	\hline
	Variable & {Coefficient} & {SE} & CI & $p$ & $df$ & $F$ \\
	
	\hline
	Intercept & 1.059 & 0.023 & $[1.015, 1.104]$ & $< 10^{-7}$ & &  \\
	Network &  &  &  & $< 10^{-7}$ & 22 & 44.734 \\
	\hspace{2 em} Karate & 0.009 & 0.029 & $[-0.048, 0.066]$ & $0.746$ && \\
	\hspace{2 em} Bat & $-0.052$ & 0.029 & $[-0.109, 0.005]$ & $0.071$ & &  \\
	\hspace{2 em} Surfer & $-0.039$ & 0.029 & $[-0.096, 0.018]$ & $0.181$&& \\
	\hspace{2 em} Elephant seal & 0.401 & 0.029 & $[0.344, 0.458]$ & $< 10^{-7}$ & & \\
	\hspace{2 em} Lizard & $-0.043$ & 0.029 & $[-0.100, 0.014]$ & $0.137$ && \\
	\hspace{2 em} Dolphin & $-0.054$ & 0.029 & $[-0.111, 0.003]$ & $0.063$ & &  \\
	\hspace{2 em} Weaver bird & 0.006 & 0.029 & $[-0.051, 0.063]$ & 0.846 && \\
	\hspace{2 em} Highschool boy & 0.025 & 0.029 & $[-0.032, 0.082]$ & $0.394$ && \\
	\hspace{2 em} Tortoise & 0.349 & 0.029 & $[0.292, 0.406]$ & $< 10^{-7}$&& \\
	\hspace{2 em} House finch & $-0.083$ & 0.029 & $[-0.140, -0.026]$ & $0.004$ && \\
	\hspace{2 em} Vole & $-0.041$ & 0.029 & $[-0.098, 0.016]$ & $0.156$ && \\
	\hspace{2 em} Nestbox & $-0.031$ & 0.029 & $[-0.087, 0.026]$ & $0.292$ && \\
	\hspace{2 em} Pira & $0.168$ & 0.029 & $[0.111, 0.265]$ & $< 10^{-7}$&& \\
	\hspace{2 em} Drug user & $0.246$ & 0.029 & $[0.189, 0.302]$ & $< 10^{-7}$ & & \\
	\hspace{2 em} Jazz & $-0.061$ & 0.029 & $[-0.118, -0.004]$ & $0.036$ && \\
	\hspace{2 em} Hall & $-0.080$ & 0.029 & $[-0.137, -0.023]$ & $0.006$ && \\
	\hspace{2 em} Netsci & $0.005$ & 0.029 & $[-0.052, 0.062]$ & $0.859$&& \\
    	\hspace{2 em} ER & $-0.050$ & 0.029 & $[-0.107, 0.007]$ & $0.087$ && \\
	\hspace{2 em} ER islands & $-0.106$ & 0.029 & $[-0.067, 0.046]$ & $0.716$ & & \\
	\hspace{2 em} BA & $0.211$ & 0.029 & $[0.154, 0.268]$ & $< 10^{-7}$ & &  \\
    \hspace{2 em} Fitness & $-0.008$ & 0.029 & $[-0.065, 0.049]$ & $0.792$ && \\
	\hspace{2 em} Configuration & $0.074$ & 0.029 & $[0.017, 0.131]$ & $0.010$&& \\
	Control parameter & & & & $<10^{-7}$ & 1 & 41.657 \\
	\hspace{2 em} $u$ & $0.055$ & 0.009 & $[0.038, 0.072]$ & $< 10^{-7}$ && \\
	Stress and noise & & & & $0.005$ & 1 & 7.724 \\
	\hspace{2 em} heterogeneous & $-0.024$ & 0.009 & $[-0.041, -0.007]$ & $0.005$ && \\ 
	Method  & & & & $<10^{-7}$ & 3 & 90.487 \\
	\hspace{2 em} minor & $-0.094$ & 0.012 & $[-0.118, -0.071]$ & $< 10^{-7}$ && \\
	\hspace{2 em} row & $-0.171$ & 0.012 & $[-0.195, -0.147]$ & $< 10^{-7}$ && \\
	\hspace{2 em} random & $-0.172$ & 0.012 & $[-0.196, -0.148]$ & $< 10^{-7}$ && \\
	\hline
\end{longtable}

\newpage

\begin{longtable}{|l r >{\centering\arraybackslash}p{0.06\textwidth} >{\centering\arraybackslash}p{0.18\textwidth} > {\centering\arraybackslash}p{0.12\textwidth} > {\centering\arraybackslash}p{0.03\textwidth} > {\centering\arraybackslash}p{0.10\textwidth}|}
	\caption{Regression results for the SIS dynamics when the dependent variable is $\tau$. We have obtained $R^2 = 0.767$ and $F = 4,490$.}
	\label{SIS tau}
	\endfirsthead
	\endhead
	\endfoot
	\hline
	Variable & {Coefficient} & {SE} & CI & $p$ & $df$ & $F$ \\
	
	\hline
	Intercept & 0.761 & 0.004 & $[0.753, 0.769]$ & $< 10^{-7}$ & &  \\
	Network &  &  &  & $< 10^{-7}$ & 22 & 516.870 \\
	\hspace{2 em} Karate & 0.115 & 0.005 & $[0.105, 0.125]$ & $< 10^{-7}$ && \\
	\hspace{2 em} Bat & $0.119$ & 0.005 & $[0.108, 0.129]$ & $< 10^{-7}$ & &  \\
	\hspace{2 em} Surfer & $0.144$ & 0.005 & $[0.134, 0.154]$ & $< 10^{-7}$&& \\
	\hspace{2 em} Elephant seal & $-0.051$ & 0.005 & $[-0.061, -0.041]$ & $< 10^{-7}$ & & \\
	\hspace{2 em} Lizard & $0.088$ & 0.005 & $[0.078, 0.098]$ & $< 10^{-7}$ && \\
	\hspace{2 em} Dolphin & $0.024$ & 0.005 & $[0.014, 0.034]$ & $3.538 \times 10^{-6}$ & &  \\
	\hspace{2 em} Weaver bird & $0.010$ & 0.005 & $[-0.000, 0.020]$ & $0.052$ && \\
	\hspace{2 em} Highschool boy & 0.080 & 0.005 & $[0.070, 0.090]$ & $< 10^{-7}$ && \\
	\hspace{2 em} Tortoise & $-0.120$ & 0.005 & $[-0.130, -0.110]$ & $< 10^{-7}$&& \\
	\hspace{2 em} House finch & $0.014$ & 0.005 & $[0.004, 0.025]$ & $0.006$ && \\
	\hspace{2 em} Vole & $-0.025$ & 0.005 & $[-0.035, -0.015]$ & $1.290 \times 10^{-6}$&& \\
	\hspace{2 em} Nestbox & $0.047$ & 0.005 & $[0.037, 0.057]$ & $< 10^{-7}$ && \\
	\hspace{2 em} Pira & $-0.141$ & 0.005 & $[-0.151, -0.130]$ & $< 10^{-7}$&& \\
	\hspace{2 em} Drug user & $-0.174$ & 0.005 & $[-0.184, -0.164]$ & $< 10^{-7}$ & & \\
	\hspace{2 em} Jazz & $-0.012$ & 0.005 & $[-0.022, -0.001]$ & $0.025$ && \\
	\hspace{2 em} Hall & $0.034$ & 0.005 & $[0.024, 0.044]$ & $< 10^{-7}$ && \\
	\hspace{2 em} Netsci & $-0.109$ & 0.005 & $[-0.119, -0.099]$ & $< 10^{-7}$&& \\
    	\hspace{2 em} ER & $-0.033$ & 0.005 & $[-0.043, -0.023]$ & $< 10^{-7}$ && \\
	\hspace{2 em} ER islands & $-0.008$ & 0.005 & $[-0.018, 0.003]$ & $0.142$ & & \\
	\hspace{2 em} BA & $0.008$ & 0.005 & $[-0.002, 0.019]$ & $0.105$ & &  \\
    	\hspace{2 em} Fitness & $-0.051$ & 0.005 & $[-0.061, -0.041]$ & $< 10^{-7}$ && \\
	\hspace{2 em} Configuration & $-0.057$ & 0.005 & $[-0.067, -0.047]$ & $< 10^{-7}$&& \\
	Stress and noise & & & & $<10^{-7}$ & 1 & 37.382 \\
	\hspace{2 em} heterogeneous & $-0.009$ & 0.001 & $[-0.012, -0.006]$ & $< 10^{-7}$ && \\ 
	Simulation direction & & & & $< 10^{-7}$ & 1 & 5,374.346 \\
	\hspace{2 em} descending & $-0.112$ & 0.001 & $[-0.115, -0.109]$ & $< 10^{-7}$ && \\
	Method  & & & & $<10^{-7}$ & 3 & 34,818.500 \\
	\hspace{2 em} minor & $-0.124$ & 0.002 & $[-0.129, -0.120]$ & $< 10^{-7}$ && \\
	\hspace{2 em} row & $-0.555$ & 0.002 & $[-0.559, -0.550]$ & $< 10^{-7}$ && \\
	\hspace{2 em} random & $-0.542$ & 0.002 & $[-0.546, -0.537]$ & $< 10^{-7}$ && \\
	\hline
\end{longtable}

\newpage

\begin{longtable}{|l r >{\centering\arraybackslash}p{0.06\textwidth} >{\centering\arraybackslash}p{0.18\textwidth} > {\centering\arraybackslash}p{0.12\textwidth} > {\centering\arraybackslash}p{0.03\textwidth} > {\centering\arraybackslash}p{0.10\textwidth}|}
	\caption{Regression results for the SIS dynamics when the dependent variable is $d$. We have obtained $R^2 = 0.656$ and $F = 2,602$.}
	\label{SIS d}
	\endfirsthead
	\endhead
	\endfoot
	\hline
	Variable & {Coefficient} & {SE} & CI & $p$ & $df$ & $F$ \\
	
	\hline
	Intercept & 5.905 & 0.039 & $[5.828, 5.981]$ & $< 10^{-7}$ & &  \\
	Network &  &  &  & $< 10^{-7}$ & 22 & 691.780 \\
	\hspace{2 em} Karate & 0.560 & 0.050 & $[0.462, 0.658]$ & $< 10^{-7}$ && \\
	\hspace{2 em} Bat & $1.844$ & 0.050 & $[1.746, 1.942]$ & $< 10^{-7}$ & &  \\
	\hspace{2 em} Surfer & $2.128$ & 0.050 & $[2.030, 2.226]$ & $< 10^{-7}$&& \\
	\hspace{2 em} Elephant seal & $-0.581$ & 0.050 & $[-0.679, -0.483]$ & $< 10^{-7}$ & & \\
	\hspace{2 em} Lizard & $0.833$ & 0.050 & $[0.735, 0.931]$ & $< 10^{-7}$ && \\
	\hspace{2 em} Dolphin & $-0.068$ & 0.050 & $[-0.166, 0.030]$ & $0.175$ & &  \\
	\hspace{2 em} Weaver bird & $-0.160$ & 0.050 & $[-0.258, -0.062]$ & $0.001$ && \\
	\hspace{2 em} Highschool boy & 0.726 & 0.050 & $[0.628, 0.823]$ & $< 10^{-7}$ && \\
	\hspace{2 em} Tortoise & $-0.825$ & 0.050 & $[-0.922, -0.727]$ & $< 10^{-7}$&& \\
	\hspace{2 em} House finch & $0.260$ & 0.050 & $[0.163, 0.358]$ & $1.818 \times 10^{-7}$ && \\
	\hspace{2 em} Vole & $-0.602$ & 0.050 & $[-0.700, -0.504]$ & $< 10^{-7}$&& \\
	\hspace{2 em} Nestbox & $1.923$ & 0.050 & $[1.826, 2.021]$ & $< 10^{-7}$ && \\
	\hspace{2 em} Pira & $-1.095$ & 0.050 & $[-1.193, -0.997]$ & $< 10^{-7}$&& \\
	\hspace{2 em} Drug user & $-1.156$ & 0.050 & $[-1.254, -1.059]$ & $< 10^{-7}$ & & \\
	\hspace{2 em} Jazz & $0.201$ & 0.050 & $[0.103, 0.299]$ & $5.560 \times 10^{-5}$ && \\
	\hspace{2 em} Hall & $0.605$ & 0.050 & $[0.507, 0.703]$ & $< 10^{-7}$ && \\
	\hspace{2 em} Netsci & $-0.989$ & 0.050 & $[-1.086, -0.891]$ & $< 10^{-7}$&& \\
    	\hspace{2 em} ER & $-0.341$ & 0.050 & $[-0.439, -0.243]$ & $< 10^{-7}$ && \\
	\hspace{2 em} ER islands & $-0.093$ & 0.050 & $[-0.191, 0.005]$ & $0.063$ & & \\
	\hspace{2 em} BA & $-0.248$ & 0.050 & $[-0.346, -0.150]$ & $6.720 \times 10^{-7}$ & &  \\
    	\hspace{2 em} Fitness & $-0.495$ & 0.050 & $[-0.593, -0.397]$ & $< 10^{-7}$ && \\
	\hspace{2 em} Configuration & $-0.384$ & 0.050 & $[-0.482, -0.287]$ & $< 10^{-7}$&& \\
	Stress and noise & & & & $<10^{-7}$ & 1 & 206.287 \\
	\hspace{2 em} heterogeneous & $-0.211$ & 0.015 & $[-0.240, -0.183]$ & $< 10^{-7}$ && \\ 
	Simulation direction & & & & $< 10^{-7}$ & 1 & 7,181.760 \\
	\hspace{2 em} descending & $-1.247$ & 0.015 & $[-1.276, -1.218]$ & $< 10^{-7}$ && \\
	Method  & & & & $<10^{-7}$ & 3 & 15,880.615 \\
	\hspace{2 em} minor & $-2.092$ & 0.021 & $[-2.133, -2.051]$ & $< 10^{-7}$ && \\
	\hspace{2 em} row & $-3.954$ & 0.021 & $[-3.994, -3.913]$ & $< 10^{-7}$ && \\
	\hspace{2 em} random & $-3.840$ & 0.021 & $[-3.880, -3.799]$ & $< 10^{-7}$ && \\
	\hline
\end{longtable}

\newpage

\bibliographystyle{unsrt} 
\bibliography{mybibliography}

\end{document}